\documentclass[12pt]{article}
\usepackage[dvips]{graphicx}
\begin{document}

\title{
Explicit two cycle model in investigation of
stochastic effects
in diffusion regime of metastable phase decay}
\author{Victor Kurasov}


\maketitle

\begin{abstract}
The theory for
manifestation of
stochastic appearance of
embryos  in the global
decay of metastable
phase has been
constructed. The regime
of droplets growth is
supposed to be both free
molecular one and diffusion one. The
deviation for a mean
droplets number from the
value predicted by the
theory based on
averaged characteristics
have been calculated.
The value of dispersion
for the distribution of
the total droplets
number in particular
attempt has been also
calculated analytically.
Comparison with results
of numerical simulation
has been given and the
correspondence between
simulation and analytical
approximate results is
rather good.

\end{abstract}

The aim of this paper is to give
analysis of decay in
diffusion regime of the droplets
growth in frames of the method
given in \cite{Koll} and corrected
in \cite{statiae}. This
method is based on the following approximation
\begin{itemize}
\item
Until the
half of the nucleation period (in the
time scale) the
supersaturation remains unperturbed and the
droplets appear
independently.
\item
In the second
half of the nucleation period the droplets
appear under
the supersaturation changed by the vapor
consumption by the droplets
appeared in  the first part of
the nucleation period.
\end{itemize}
The part of the droplets size
spectrum which corresponds to
the droplets appeared during the first part of the
nucleation period will be
called as the first part of
spectrum, the remaining
part will be the second part of
spectrum.

Here this model will be generalized by means of the
following changes:
\begin{itemize}
\item
The duration of the
first part of nucleation period will be
not exactly the first half
but it will be regulated by some
parameter $p$ instead of $1/2$.
\item
The regime of growth will be an
arbitrary power law of growth
(the power type is adopted
only to present concrete
formulas, it isn't  principal here).
\item
It is supposed that the
distribution of droplets in the
first part of spectrum can differ
from the distribution
$$
P_0 (N) \sim \exp(\frac{(N-\bar{N})^2}{2 \bar{N}})
$$
of $N$ independently appeared
droplets during the period
of time corresponding to the mean value $\bar{N}$ of
appeared droplets.

Now instead of the previous distribution
the distribution
with renormalized dispersion
$$
P_r (N) \sim
\exp(\frac{(N-\bar{N})^2}{2 \psi \bar{N}})
$$
will be used. Here $\psi$ is parameter of
renomalization. This change is
initiated by the using of
the property of self similarity of
spectrums in the first
iteration
(see \cite{varios}) used for investigation of
stochastic effects in \cite{statiae}.
\end{itemize}

The structure of consideration
will be the following
\begin{itemize}
\item
At first we shall present the
formal generalized model
\item
Then the parameter $p$ of the boundary
between the first and
the second parts will be calculated.
\item
In the last part the effect
of the "growing volumes of
interaction" will be described
and it will be shown how to
include this effect into already presented scheme.
\item
In Appendix the theory
for the free molecular
regime of growth will be
presented. This theory
was given in
\cite{statiae} but now
some modifications and
improvements have been
made. So, a new variant
can be found in
appendix.

\end{itemize}

The main object of investigation will be the
calculation of
dispersion of the droplets distribution since
the mean
value will be very close to the characteristic
calculated
in frames of the theory based on the averaged
characteristics (TAC),
all arguments given for the free molecular
regime
of the droplets growth  remain here practically the
same \cite{statiae}, the only
principally new feature  is to use the theory with
explicit profiles of density around the growing
droplets
\cite{PhysicaA}.

The smallness of fluctuations
which will be shown here plays
a very important role in
transition from the free molecular
kinetics to the diffusion kinetics. Really, such
transition was shown in \cite{PhysicaA} only
in general features on the level of TAC
and the possible giant fluctuations initiated by
stochastic appearance
of droplets would destroy this transition in
frames of the known
approaches. Fortunately the giant
fluctuations don't appear and
the justification of their
absence lies in the moderate
value of dispersion calculated
here.

\section{Calculation of dispersion}

Here we shall analyze the situation with
diffusion regime of metastable phase consumption.
At it has been already noticed this situation is
characterized by the growing volumes of
interaction which produces additional
difficulties. The growing
volume of interaction means
that the fixed point in the
volume will be under influence of
vapor consumers which appeared in a volume
$V$ with a center at this point and with a
radius $\sqrt{4Dt}$
where $t$ is a time counted from the
consumer appearance
up to a current moment
and $D$ is a diffusion coefficient.
Certainly the last value
can be regarded only as the rough estimate,
the precise expression can be obtained only of
the base of precise profiles
of  a substance gap \cite{PhysicaA}.

At first we briefly follow the way to derive
the estimate
for dispersion proposed
in \cite{Koll}   to use it for diffusion
regime of metastable phase consumption.
As we have noticed the formalism of a volume of
interaction is a rather rough estimate (precise
theory has to be based on a form of substance gap
from \cite{PhysicaA}).
Thus, the accuracy here is limited.
So, in frames of adopted accuracy
we shall use here only some
rough way to estimate dispersion of a distribution
 and can use formalism
from \cite{Koll} with
appropriate comments and modifications.
This way  is useful because it
allows to understand
formulas from \cite{Koll} which were widely
used in above consideration.

Here we suppose that the rate of
embryos growth is given by
$$
\frac{d\nu}{dt} \sim \nu^s
$$
where $\nu$ is the number of molecules
inside the embryo and $s$ is a power
parameter. We suppose that $s$ takes values
which are essentially greater than $0$ ($s$ isn't
too small in comparison with $1$) and
essentially less than $1$ ($1-s$ isn't too small in
comparison with $1$). For the free molecular
regime of metastable phase consumption
$s=2/3$, for the diffusion regime $s=1/3$.

Then the characteristic
$\rho$ of a droplet defined as
$$
\rho =\nu^{(1/s)+1}
$$
grows with velocity independent on it's value
$$
d \rho / dt \sim 1
$$

The period of
nucleation is divided in many
elementary intervals
with a length
$\Delta$ (or $\Delta_i$) in a $\rho$-scale,
the average total number of interval in the
whole nucleation
period is $M$, the number of embryos $N_i$
appeared during every
elementary interval is
supposed to be big $N_i \gg 1$.

The fact
that the total number
of intervals is $M$ and their length is $\Delta$
(in estimates we can put it one and the same for all
intervals, at least we can
take the smallest
length and attribute it to all intervals) means that
$$\Gamma \bar{f} (M \Delta)^{(\frac{1}{1-s}+1)}
 \sim {(\frac{1}{1-s}+1)} \zeta(0)
 \ , $$
  where
$\zeta(0)$ is the initial
supersaturation, $\bar{f}$ is the
initial average amplitude of embryos
size spectrum \cite{PhysRevE94}, $\Gamma$ is
parameter from TAC \cite{PhysRevE94}.
Namely the last
relation can be derived
from the first iteration  in TAC.
It will be used to express $\bar{f}$
through $M, \Delta$.

During every interval the increase
$\delta \rho$  of the
value $\rho$ of already existing
droplet will be
$$
\delta \rho \sim  \Delta \zeta (0) / \tau
$$
with some characteristic time  $\tau$ which will be
dropped out
in final formulas.

Denote by $N_i$ the number
of embryos appeared at $i$-th interval.
The value $\bar{N}_i$ is the
average $N_i$, $P_i(N_i)$ is the
density distribution on $N_i$.

Denote by $N^{(i)}$ the
number of embryos appeared at the
first $i$-th intervals $\Delta$.
The value $\bar{N}^{(i)}$
is the average $N^{(i)}$, $P^{(i)}(N^{(i)})$ is
the density distribution of $N^{(i)}$.
It is important to stress that $\bar{N}^{(i)}$ isn't equal
to $\tilde{N}^{(i)}$ which is the
corresponding value
completely
calculated in
frames of TAC (stochastically appeared
droplets in the previous moments of
time taken into account).

At the first $P$ intervals (i.e.
at the first part of nucleation
period) the metastable substance
consumption is negligible.
Here $P = p M$ and $p$ is parameter of constructions.
Then
for $k<P$ the distribution
$P^{(k)}$ of the number of
droplets $N^{(k)}$ appeared in the
first $k$ intervals is
$$
P^{(k)} (N^{(k)}) \sim
\exp(-
\frac{(N^{(k)} - \bar{N}^{(k)})^2 }
{2 \psi \bar{N}^{(k)}} )
$$
and it is
is a standard Gaussian
distribution of $k$
independent
subsystems\footnote{Here and later
without any notation we
don't  keep the preexponential factors
before  Gaussian distributions, they can
be easily reconstructed by integration.}.

We introduce here the parameter of
renormalization $\psi$
to be able to fulfill
calculations suggested in \cite{statiae}
based on the approximate similarity of
nucleation conditions \cite{varios}.

For
$k>P$ the expression for $P^{(k)}$ will be
formally the following
$$
P^{(k)}(N^{(k)}) =
\sum_{N_1, ..... , N_{P}}
\prod_{i=1}^{P} \hat{P}_i (N_i)  \Omega
$$
where
$$
\hat{P}_i (N_i) \sim
\exp(-\frac{(N_i - \bar{N}_i)^2}{2 \psi N_i })
$$
$$
\Omega = \sum_{N_{P+1}, .... , N_{k-1}}
 \prod_{j=P+1}^{k-1} \hat{P}_j (N_j)
\hat{P}_k(N^{(k)}-\sum_{l=1}^{k-1} N_l )
$$
Here the function $\Omega$ is extracted because
it corresponds to the second part of spectrum.
The function $\Omega$ is superposition of
independent subsystems and can be
calculated explicitly
$$
\Omega \sim
\exp(-\frac{ (
N^{(k)} - \sum_{i=1}^{P} N_i -
\sum_{j=P+1}^{k} \bar{N}_j
)^2}
{2\sum_{j=P+1}^{k} \bar{N}_j })
$$

The problem is that
the values $\bar{N}_j$ standing in
the last expression are
unknown now. They are
functions of all $N_i, \ \ i=1,..,P$
(since the two parts model is used).
So, the distributions  from the first
group $i \leq P$ have the
influence on $\Omega$.
This influence is given through
the values $\bar{N}_j $. Now we
shall get an expression for
this value.
At first we recall the
exponential approximation which can
be found elsewhere (see \cite{PhysRevE94})
$$
\bar{N}_j =
\bar{N}_1 \exp(-\frac{\Gamma}{\zeta_0} G_j )
$$
Here $G_{j}$ is the number of
molecules in a liquid phase
at $j$-th interval.
For this value we can write
$$
G_j = \sum_{i=1}^j N_{i} \rho_i^{(j)\ \frac{1}{1-s}}
$$
where
$\rho_i^{(j)}$ is the "size" $\rho$ of a
droplet appeared at
$i$-th interval at the moment
corresponding to the $j$-th
interval.
Then
$$
G_j = \sum_{i=1}^j N_{i} ((\Delta \zeta_0 /\tau) j-
(\Delta \zeta_0 / \tau)
i)^{\frac{1}{1-s}} =
 \sum_{i=1}^j N_{i} (j-i)^{\frac{1}{1-s}}
 (\Delta \zeta_0 / \tau) ^{\frac{1}{1-s}}
$$
and
$$
\bar{N}_j =
\bar{N}_1 \exp(-\frac{\Gamma}{\zeta_0}
(\Delta \zeta_0 / \tau) ^{\frac{1}{1-s}}
 \sum_{i=1}^j N_{i} (j-i)^{\frac{1}{1-s}}
 )
$$
Now we have to get another expression for
$$
K =
\frac{\Gamma}{\zeta_0}
(\Delta \zeta_0 / \tau) ^{\frac{1}{1-s}}
$$
which appeared in the last relation.
One has to clarify the
meaning of $M$ as the total number
of intervals. Certainly, it is no more than a
conditional definition to say that the end of
nucleation is the moment when the averaged
rate of nucleation falls $e$ times in
comparison with initial value.
Then the corresponding equation will be
$$
\frac{\Gamma}{\zeta_0} G_M = 1
$$
We have to
calculate $G_M$ in some approximation. Let it be
the ideal approximation, i.e. the first
iteration \cite{PhysRevE94}.
Then
$$
K \sum_{i=1}^M \bar{N}_1 (M - i)^{1/(1-s)} =1
$$
or
$$
K \bar{N}_1 \sum_{i=1}^M  (M - i)^{1/(1-s)} =1
$$
For $M \gg 1$ summation
can be replaced by integration
which gives
$$
\sum_{i=1}^M  (M - i)^{1/(1-s)}
= \int_0^M (M-x)^{1/(1-s)} dx
=  \frac{M^{1/(1-s)+1}}{\frac{1}{1-s} +1}
$$
and
\begin{equation}
\label{M}
K= ( \bar{N}_1 \frac{M^{1/(1-s)+1}}{\frac{1}{1-s} +1}
)^{-1}
\end{equation}
Then
$$
\bar{N}_j =
\bar{N}_1
\exp(-\frac{1}{\bar{N}_1 M^{(\frac{1}{1-s}+1)}}
\sum _{i=1}^{j}
N_i {(\frac{1}{1-s}+1)} (j-i)^{\frac{1}{1-s}} )
$$
Here
$$
\frac{1}{\bar{N}_1 M^{(\frac{1}{1-s}+1)}}
$$
is simply the scaling factor.
One can argue whether the style of
calculation of $G_M$ is
appropriate. Really we used
another approximation which
differs from the two cycle model which we
are going to use.
But there is absolutely no sense how to
calculate the last
value. One can simply say that our model
is to take the
first part as $P=pM$ where
the definition of $M$ is
given by
(\ref{M}).
The problem is how to choose $p$.

We continue to simplify
expression for $\bar{N_j}, \  j>P$.
We recall that the droplets from
the first part only
consume vapor. Then we  come to
\begin{equation} \label{RR}
\bar{N}_j =
\bar{N}_1
\exp(-\frac{1}{\bar{N}_1 M^{(\frac{1}{1-s}+1)}}
\sum _{i=1}^{P}
N_i {(\frac{1}{1-s}+1)} (j-i)^{\frac{1}{1-s}} )
\end{equation}

Now we shall simplify the last formula. We make the
following transformations:

- we add and subtract $\bar{N}_1$,
which is the ideal mean
value of droplets appeared during the
elementary interval.
Then
$$\bar{N}_j
=
\bar{N}_1 \exp(-\frac{1}{ M^{(\frac{1}{1-s}+1)}}
\sum _{i=1}^{P}
(1+\frac{N_i-\bar{N}_1}{\bar{N}_1})
 {(\frac{1}{1-s}+1)}(j-i)^{\frac{1}{1-s}} )
$$
and
\begin{eqnarray}
\bar{N}_j
=
\bar{N}_1 \exp(-\frac{1}{ M^{(\frac{1}{1-s}+1)}}
\sum _{i=1}^{P}
 {(\frac{1}{1-s}+1)}(j-i)^{\frac{1}{1-s}}
 -
 \nonumber
 \\
 \nonumber
\frac{1}{ M^{(\frac{1}{1-s}+1)}}
\sum _{i=1}^{P}
\frac{N_i-\bar{N}_1}{\bar{N}_1}
 {(\frac{1}{1-s}+1)}(j-i)^{\frac{1}{1-s}}
 )
\end{eqnarray}
We can calculate $\sum _{i=1}^{P}
 {(\frac{1}{1-s}+1)}(j-i)^{\frac{1}{1-s}}$.
 For $M\gg 1$
 $$
\sum _{i=1}^{P}
 {(\frac{1}{1-s}+1)}(j-i)^{\frac{1}{1-s}} =
 {(\frac{1}{1-s}+1)}
 \int_0^{P} (j - x)^{\frac{1}{1-s}} dx
$$
and
since we suppose $1/3 < s< 2/3 $ we have
$ 1/(1-s) \gg 1 $ and
$$
 \int_0^{P} (j - x)^{\frac{1}{1-s}} dx
 =
 \int_0^{j} (j - x)^{\frac{1}{1-s}} dx
 =
(\frac{1}{1-s} +1)^{-1} j^{\frac{1}{1-s} +1}
$$
for
$j$ which aren't too big in
comparison with $P$ (namely
the parameter $p$ will be chosen
in such a way that $P$ can
not be too small in comparison
with $M$ and certainly $j$
can not be essentially greater than $M$).

Then
\begin{eqnarray}
\bar{N}_j
=
\bar{N}_1 \exp(-\frac{1}{ M^{(\frac{1}{1-s}+1)}}
 j^{\frac{1}{1-s}+1}
 -
 \nonumber
 \\
 \nonumber
 \frac{1}{ M^{(\frac{1}{1-s}+1)}}
\sum _{i=1}^{P}
\frac{N_i-\bar{N}_1}{\bar{N}_1}
 {(\frac{1}{1-s}+1)}(j-i)^{\frac{1}{1-s}}
 )
\end{eqnarray}
and
$$
\bar{N}_j
=
\bar{N}_1 \exp(-\frac{1}{ M^{(\frac{1}{1-s}+1)}}
 j^{\frac{1}{1-s}+1})
 \exp( -
 \frac{1}{ M^{(\frac{1}{1-s}+1)}}
\sum _{i=1}^{P}
\frac{N_i-\bar{N}_1}{\bar{N}_1}
 {(\frac{1}{1-s}+1)}(j-i)^{\frac{1}{1-s}}
 )
$$
Now we linearize the last exponent
over
$$
 \frac{1}{ M^{(\frac{1}{1-s}+1)}}
\sum _{i=1}^{P}
\frac{N_i-\bar{N}_1}{\bar{N}_1}
 {(\frac{1}{1-s}+1)}(j-i)^{\frac{1}{1-s}}
 $$
 This linearization is possible and it
 differs from linearization over
$$
 \frac{1}{ M^{(\frac{1}{1-s}+1)}}
\frac{N_i-\bar{N}_1}{\bar{N}_1}
 {(\frac{1}{1-s}+1)}(j-i)^{\frac{1}{1-s}}
 $$
 made in \cite{Koll}.
 The  linearization in \cite{Koll}
 since we study fluctuations namely
 in $N_i$ is very
 doubtful.
 In our approach we can see the compensation of
 fluctuation during the
 period comparable with the whole
nucleation period and, thus,
the linearization is possible.

It is necessary to stress that
after the linearization one
can not pretend on the derivation of
deviation of the mean
value of droplets from the
value prescribed by TAC.
Contrary to \cite{Koll} we don't
pretend on this but
here we are
interested only in dispersion.
To prove the smallness of
deviation of the mean value of
total droplets from the
value given by TAC one can use
other approaches (see
\cite{statiae}.

Here we directly use
${(\frac{1}{1-s}+1)}(j-i)^{\frac{1}{1-s}}$
as the quantity of substance in droplets
of a given size
(the relative deviation of the
size during the elementary
interval is small)
instead of inappropriate integration
over constant distribution (it is stochastic and
isn't constant)
in \cite{Koll}.

In the zero approximation we get
$$
\bar{N}_j =
\bar{N}_1\exp(-\frac{j^{(\frac{1}{1-s}+1)}}
{M^{(\frac{1}{1-s}+1)}})
$$

In the first approximation
$$
\bar{N}_j = \bar{N}_1
\exp(-\frac{j^{\frac{1}{1-s}
+1}}{M^{\frac{1}{1-s} +1}})
(1- \frac{1}{M^{\frac{1}{1-s} +1}}
\sum_{i=1}^P
\frac{N_i - \bar{N}_1}{\bar{N}_1}
(1+\frac{1}{1-s})
(j-i)^{\frac{1}{1-s}}
$$

Now we know the influence
of embryos
of big sizes and can directly
analyze the numerator of Gaussian distribution for
$\Omega$.
We have
$$
( N^{(k)} -
\sum_{i=1}^P N_i - \sum_{j=P+1}^k \bar{N}_j )
=
( N^{(k)} - \sum_{i=1}^P (N_i -\bar{N}_1)
- \sum_{i=1}^P \bar{N}_1
 - \sum_{j=P+1}^k \bar{N}_j )
 $$
 \begin{eqnarray}
( N^{(k)} - \sum_{i=1}^P (N_i -\bar{N}_1)
- \sum_{i=1}^P \bar{N}_1
 - \sum_{j=P+1}^k \bar{N}_j )
 =
 \nonumber
 \\
 \nonumber
( N^{(k)} - \sum_{i=1}^P (N_i -\bar{N}_1)
+ P \bar{N}_1
 - \sum_{j=P+1}^k \bar{N}_j )
 \end{eqnarray}
Now we shall derive formula for
$\sum_{j=P+1}^k \bar{N}_j$.
We start with (\ref{RR})
where $\bar{N}_0 \equiv \bar{N}_1$
and get
$$
\sum_{j=P+1}^k \bar{N}_j =
\bar{N}_0
\sum_{j=P+1}^k  \exp(-\frac{1}{M^{1/(1-s)+1}}
\frac{1}{\bar{N}_0}
\sum_{i=1}^P (\frac{1}{1-s} +1) (j-i)^{1/(1-s)} N_i )
$$
Since $M \gg 1$, $P \sim M$
$$
\sum_{i=1}^P (\frac{1}{1-s} +
1)(j-i)^{1/(1-s)} /M^{1/(1-s)+1}
\approx
j^{1/(1-s)+1} /M^{1/(1-s)+1}
$$
So,
one can come to
\begin{eqnarray}
\sum_{j=P+1}^k \bar{N}_j =
\bar{N}_0
\sum_{j=P+1}^k
\exp(-(j/M)^{M^{1/(1-s)+1}} )
\exp(-\frac{1}{M^{1/(1-s)+1}}
\nonumber
\\
\nonumber
\frac{1}{\bar{N}_0}
\sum_{i=1}^P (\frac{1}{1-s} +
1) (j-i)^{1/(1-s)} N_i )
\exp(
\sum_{i=1}^P (\frac{1}{1-s} +
1)(j-i)^{1/(1-s)} /M^{1/(1-s)+1}
)
\end{eqnarray}

Then
\begin{eqnarray}
\sum_{j=P+1}^k \bar{N}_j =
\bar{N}_0
\sum_{j=P+1}^k
\exp(-(j/M)^{M^{1/(1-s)+1}} )
\nonumber
\\
\nonumber
\exp(-\frac{1}{M^{1/(1-s)+1}}
\sum_{i=1}^P (\frac{1}{1-s} +1) (j-i)^{1/(1-s)}
\frac{(N_i  - \bar{N}_0)}{\bar{N}_0}
)
\end{eqnarray}

Now one can fulfill linearization of exponent over
$$
\frac{1}{M^{1/(1-s)+1}}
\sum_{i=1}^P (\frac{1}{1-s} +1) (j-i)^{1/(1-s)}
\frac{(N_i  - \bar{N}_0)}{\bar{N}_0}
$$
which leads to
\begin{eqnarray}
\sum_{j=P+1}^k \bar{N}_j =
\bar{N}_0
\sum_{j=P+1}^k
\exp(-(j/M)^{M^{1/(1-s)+1}} )
\nonumber
\\
\nonumber
(1-\frac{1}{M^{1/(1-s)+1}}
\sum_{i=1}^P (\frac{1}{1-s} +1) (j-i)^{1/(1-s)}
\frac{(N_i  - \bar{N}_0)}{\bar{N}_0}
)
\end{eqnarray}
The formula is ready.

Then
\begin{eqnarray}
( N^{(k)} - \sum_{i=1}^P (N_i -\bar{N}_1)
- P \bar{N}_1
 - \sum_{j=P+1}^k \bar{N}_j )
 =
( N^{(k)} - \sum_{i=1}^P (N_i -\bar{N}_1)
- P \bar{N}_1
 -
 \nonumber
 \\
 \nonumber
\bar{N}_0
\sum_{j=P+1}^k
\exp(-(j/M)^{1/(1-s)+1} )
(1-\frac{1}{M^{1/(1-s)+1}}
\sum_{i=1}^P (\frac{1}{1-s} +1) (j-i)^{1/(1-s)}
\frac{(N_i  - \bar{N}_0)}{\bar{N}_0}
)
 )
 \end{eqnarray}
The next transformation gives
\begin{eqnarray}
( N^{(k)} - \sum_{i=1}^P (N_i -\bar{N}_1)
- P \bar{N}_1
 -
\bar{N}_0
\sum_{j=P+1}^k
\exp(-(j/M)^{1/(1-s)+1} )
(1-\frac{1}{M^{1/(1-s)+1}}
\nonumber
\\
\sum_{i=1}^P (\frac{1}{1-s} +1) (j-i)^{1/(1-s)}
\frac{(N_i  - \bar{N}_0)}{\bar{N}_0}
)
 )
=
\nonumber
\\
\nonumber
( N^{(k)} - \sum_{i=1}^P (N_i -\bar{N}_1)
- P \bar{N}_1
 -
 \bar{N}_0
\sum_{j=P+1}^k
\exp(-(j/M)^{1/(1-s)+1} )
+
\\
\nonumber
\bar{N}_0
\sum_{j=P+1}^k
\exp(-(j/M)^{1/(1-s)+1} )
\frac{1}{M^{1/(1-s)+1}}
\sum_{i=1}^P (\frac{1}{1-s} +1) (j-i)^{1/(1-s)}
\frac{(N_i  - \bar{N}_0)}{\bar{N}_0}
 )
 \end{eqnarray}

Having noticed that
$$
\tilde{N}^{(k)} =
\bar{N}_1 (P + \sum_{j=P+1}^k
\exp(-\frac{j^{(\frac{1}{1-s}+
1)}}{M^{(\frac{1}{1-s}+1)}})
$$
is  the mean number of droplets $N^{(k)}$
prescribed in the zero approximation by TAC,
we come to
\begin{eqnarray}
( N^{(k)} - \sum_{i=1}^P (N_i -\bar{N}_1)
- P \bar{N}_1
 -
 \bar{N}_0
\sum_{j=P+1}^k
\exp(-(j/M)^{1/(1-s)+1} )
+
\nonumber
\\
  \bar{N}_0
\sum_{j=P+1}^k
\exp(-(j/M)^{1/(1-s)+1} )
\frac{1}{M^{1/(1-s)+1}}
\sum_{i=1}^P (\frac{1}{1-s} +1) (j-i)^{1/(1-s)}
\frac{(N_i  - \bar{N}_0)}{\bar{N}_0}
 )
 =
 \nonumber
 \\
 \nonumber
 ( N^{(k)} - \sum_{i=1}^P (N_i -\bar{N}_1)
- \tilde{N}^{(k)}
+   \bar{N}_0
\sum_{j=P+1}^k
\exp(-(j/M)^{1/(1-s)+1} )
\frac{1}{M^{1/(1-s)+1}}
\\
\nonumber
\sum_{i=1}^P (\frac{1}{1-s} +1) (j-i)^{1/(1-s)}
\frac{(N_i  - \bar{N}_0)}{\bar{N}_0}
 )
 \end{eqnarray}

Now we shall change the
order of summation and come to
\begin{eqnarray}
 ( N^{(k)} - \sum_{i=1}^P (N_i -\bar{N}_1)
- \tilde{N}^{(k)}
+
\nonumber
\\
\bar{N}_0
\sum_{j=P+1}^k
\exp(-(j/M)^{1/(1-s)+1} )
\frac{1}{M^{1/(1-s)+1}}
\sum_{i=1}^P (\frac{1}{1-s} +1) (j-i)^{1/(1-s)}
\frac{(N_i  - \bar{N}_0)}{\bar{N}_0}
 )
=
\nonumber
\\
\nonumber
 ( N^{(k)} - \sum_{i=1}^P (N_i -\bar{N}_1)
- \tilde{N}^{(k)}
+
\\
\nonumber
\sum_{i=1}^P
(N_i  - \bar{N}_0)
\sum_{j=P+1}^k
\exp(-(j/M)^{1/(1-s)+1} )
\frac{1}{M^{1/(1-s)+1}}
 (\frac{1}{1-s} +1) (j-i)^{1/(1-s)}
 )
\end{eqnarray}

The last transformation leads to
\begin{eqnarray}
 ( N^{(k)} - \sum_{i=1}^P (N_i -\bar{N}_1)
- \tilde{N}^{(k)}
+
\nonumber
\\
\sum_{i=1}^P
(N_i  - \bar{N}_0)
\sum_{j=P+1}^k
\exp(-(j/M)^{1/(1-s)+1} )
\frac{1}{M^{1/(1-s)+1}}
 (\frac{1}{1-s} +1) (j-i)^{1/(1-s)}
)
 =
 \nonumber
 \\
 \nonumber
 ( N^{(k)}
- \tilde{N}^{(k)}
-
\sum_{i=1}^P
(N_i  - \bar{N}_0) (1 -
\\
\nonumber
\sum_{j=P+1}^k
\exp(-(j/M)^{1/(1-s)+1} )
\frac{1}{M^{1/(1-s)+1}}
 (\frac{1}{1-s} +1) (j-i)^{1/(1-s)}
 )
)
\end{eqnarray}
or
\begin{eqnarray}
 ( N^{(k)} - \sum_{i=1}^P (N_i -\bar{N}_1)
- \tilde{N}^{(k)}
+
\nonumber
\\
\sum_{i=1}^P
(N_i  - \bar{N}_0)
\sum_{j=P+1}^k
\exp(-(j/M)^{1/(1-s)+1} )
\frac{1}{M^{1/(1-s)+1}}
 (\frac{1}{1-s} +1) (j-i)^{1/(1-s)}
)
 =
 \nonumber
 \\
 \nonumber
 ( N^{(k)}
- \tilde{N}^{(k)}
-
\sum_{i=1}^P
(N_i  - \bar{N}_0)
a_i^{(k)}
)
\end{eqnarray}
where
$$
a_i^{(k)} =
(1 -
\sum_{j=P+1}^k
\exp(-(j/M)^{1/(1-s)+1} )
\frac{1}{M^{1/(1-s)+1}}
 (\frac{1}{1-s} +1) (j-i)^{1/(1-s)}
 )
 $$
The last definition can be rewritten as
$$
a_i^{(k)} =
(1-b_i^{(k)} )
$$
where
$$
b_i^{(k)} =
\sum_{j=P+1}^k
\exp(-(j/M)^{1/(1-s)+1} )
\frac{1}{M^{1/(1-s)+1}}
 (\frac{1}{1-s} +1) (j-i)^{1/(1-s)}
$$

Now we shall
analyze denominator of Gaussian distribution for
$\Omega$.
Here we can use $\bar{N}_j$ in
the zero approximation and come to
$$
\sum_{j= P +1}^{k} \bar{N}_j  =
\tilde{N}^{(k)} - P \bar{N}_1
$$
The use of zero approximation
is based on the following
simple estimate. Really, the
last linearization is based on
the smallness mainly of the parameter
which can be simply
neglected in denominator since
$$
\sum_{j=P+1}^k \exp(-(\frac{j}{M})^{1/(1-s)+1})
$$
isn't too small.

Now we can directly go to the calculation of
$P^{(k)}
( N^{(k)} )
$
for
$k>P$.
The starting formula is
\begin{eqnarray} \label{starting}
P^{(k)} (N^{(k)}) =
\int_{-\infty}^{\infty} d (N_1 - \bar{N}_1) ...
\int_{-\infty}^{\infty} d (N_{P} - \bar{N}_P)
\prod_{i=1}^{P} \exp(-\frac{(N_i -
\bar{N}_1)^2}{2 \psi \bar{N}_1} )
\nonumber \\
\\ \nonumber
\exp(-
\frac{[
N^{(k)} - \tilde{N}^{(k)} -
\sum_{i=1}^{P}
 a_i^{(k)}
 (N_i - \bar{N}_1)
]^2}{
2 (\tilde{N}^{(k)} - P \bar{N}_1 )
})
\end{eqnarray}
where we because of $\bar{N}_i \gg 1$ have replaced
summation by integration. The limits of
integration can be infinite ones  because of
 $\bar{N}_i \gg 1$ and estimates
 predicted by Gaussian
 distribution.

To fulfill integration we have to use a simple
formula
$$
\int_{-\infty}^{\infty}
\exp(-\frac{x^2}{c})
\exp(-\frac{(lx+d)^2}{b}) dx
\sim \exp(-\frac{d^2}{b+l^2 c})
$$
We have to use this formula $P$ times with
$$
x \sim x_i = N_i - \bar{N}_1
$$
$$
c \sim c_i = 2 \psi \bar{N}_1
$$
$$
l \sim l_i =  - a_i^{(k)}
$$
$$
b \sim b_i =2 (\tilde{N}^{(k)} - P \bar{N}_1 )
$$

This
leads to
$$
P^{(k)} (N^{(k)})
\sim
\exp(-\frac{
(N^{(k)} - \tilde{N}^{(k)} )^2
}{
2 D^{(k)}
})
$$
where dispersion $D^{(k)}$
is given by
$$
D^{(k)} =
\tilde{N}^{(k)} -
(P -
\psi \sum_{i=1}^{P} a_i^{(k)^2}) \bar{N}_1
 $$

Now we can calculate dispersion at infinite $k$.
Since $M \gg 1$ we can substitute summation by
integration. The
number of interval corresponds to the
variable $z*M$, then $z=1$ corresponds to the end of
nucleation.

Since
$$
\tilde{N}^{(\infty)} =
 \bar{N}_1 (P + \sum_{j=P+1}^{\infty}
 \exp(-(\frac{j}{M})^{1/(1-s)+1} ) )
 \approx M \bar{N}_1 \alpha
 $$
 $$
  \alpha \equiv \int_0^{\infty}
 \exp(-x^{1/(1-s)+1}) dx
 $$
then
$$
D^{(\infty)} =
\tilde{N}^{(\infty)} (1- (1-\psi) \frac{P}{M\alpha}
- \frac{2 \psi}{\alpha} \frac{1}{M}
\sum_{i=1}^P b_i^{(\infty)}
+
\frac{ \psi}{\alpha} \frac{1}{M}
\sum_{i=1}^P b_i^{(\infty)^2})
$$
One can easily calculate
$b_i^{(\infty)}$:
\begin{eqnarray}
b_i^{(\infty)} =
\frac{1/(1-s)+1}{M^{1/(1-s)+1}}
\sum_{j=P}^{\infty}
\exp(-(\frac{j}{M})^{1/(1-s)+1}) (j-i)^{1/(1-s)}
\nonumber
\\
\nonumber
\rightarrow
(1/(1-s)+1)
\int_{p}^{\infty}
\exp(-x^{1/(1-s)+1}) (x-y)^{1/(1-s)}
dx
\end{eqnarray}
Also one can note that
$$
\frac{1}{M} \sum_{i=1}^{P}
\rightarrow
\int_0^p dy
$$
Then
$$
D^{(\infty)} =
\tilde{N}^{(\infty)}
(1- (1-\psi) \frac{p}{\alpha}- \frac{\Psi}{\alpha} \beta)
$$
where
$$
\beta = \beta_1 - \beta_2
$$
$$
\beta_1 = 2 (1/(1-s)+1) \int_0^p dy
\int_p^{\infty} dx
\exp(-x^{1/(1-s)+1}) (x-y)^{1/(1-s)}
$$
$$
\beta_2 =  (1/(1-s)+1)^2 \int_0^p dy
(\int_p^{\infty} dx
\exp(-x^{1/(1-s)+1}) (x-y)^{1/(1-s)})^2
$$
This  result can be directly calculated.

\section{Calculations of  parameters of the model}

To compare analytical  constructions
with the real value of
dispersion we shall present numerical simulations.

The results for the mean value of droplets
calculated in units of the
droplets number
calculated in TAC as function of the droplets
number calculated in TAC are presented in
figure 1.

\begin{figure}[hgh]

\includegraphics[angle=270,totalheight=10cm]{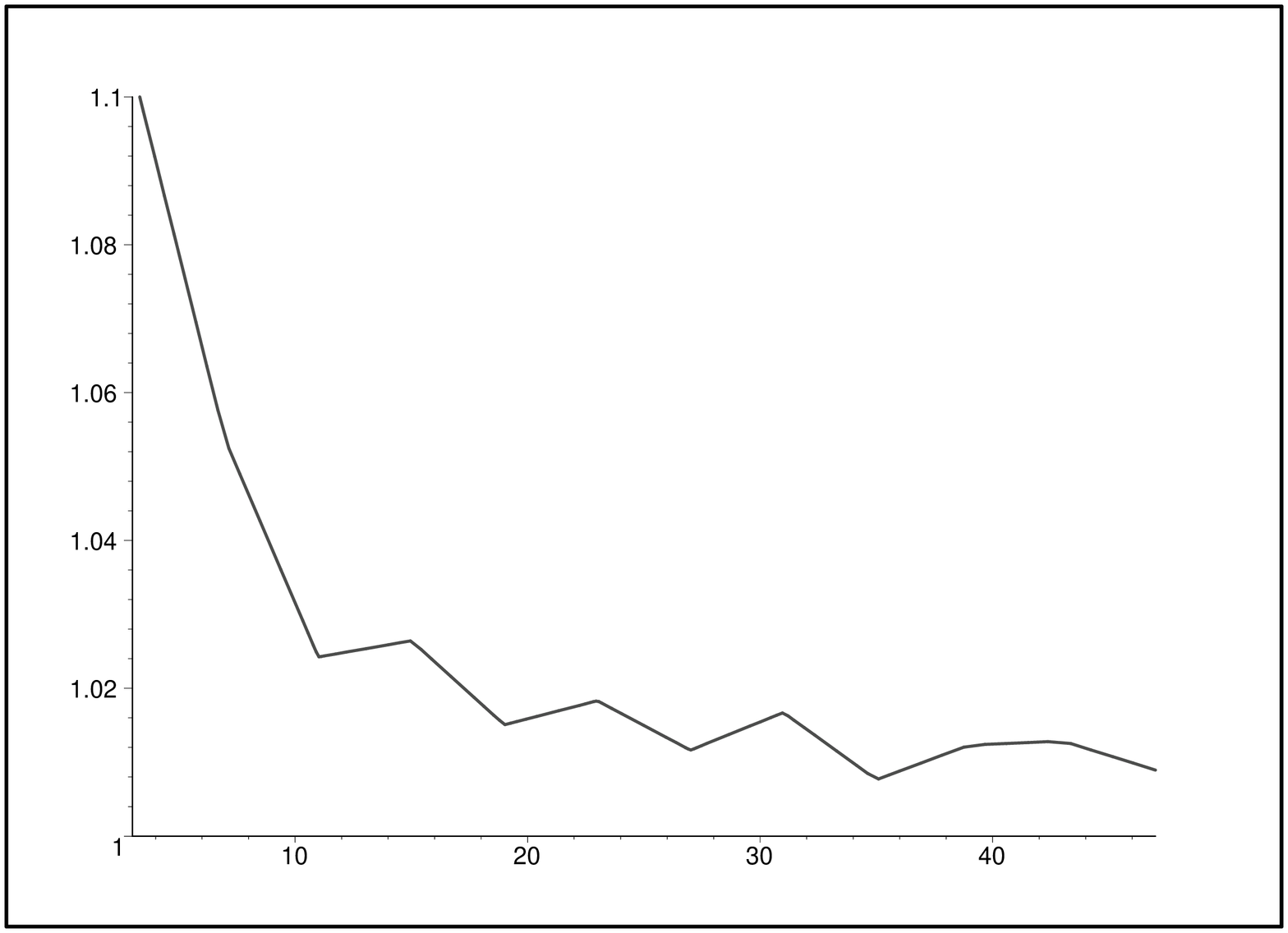}

\begin{caption}
{Mean value of the droplets number}
\end{caption}

\end{figure}

The relative
dispersion as a function of the
mean number of droplets
is shown in figure 2.


\begin{figure}[hgh]

\includegraphics[angle=270,totalheight=10cm]{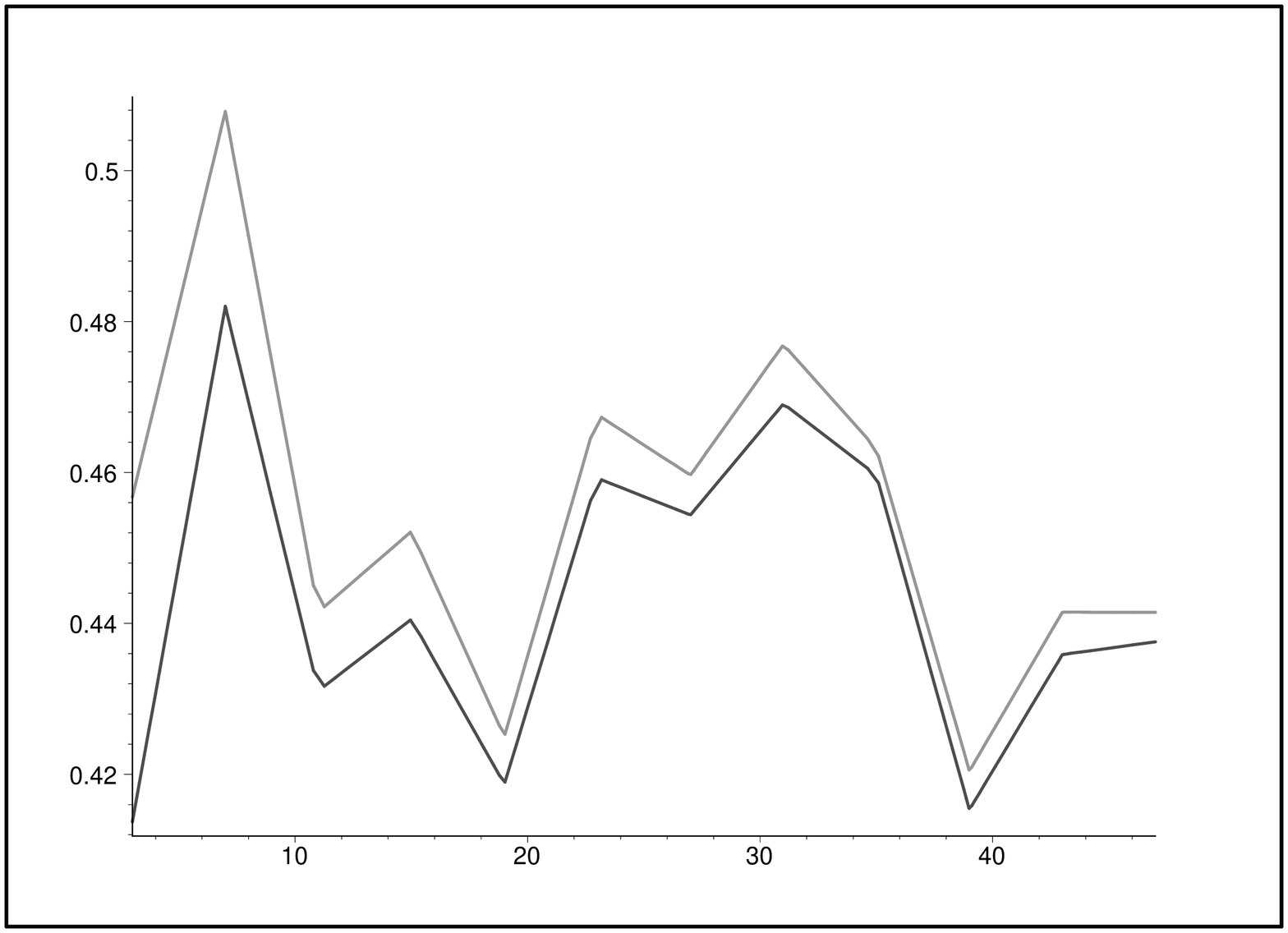}

\begin{caption}
{Dispersion}
\end{caption}

\end{figure}

Now we shall perform calculations.
We study equation
$$
g(x) = \frac{5}{2} \int_0^z \exp(-g(x)) (z-x)^{3/2} dx
$$
Namely the
coefficient $\frac{5}{2}$ corresponds to the
final formulas in the previous section.

At first we calculate the form of
spectrum. It is shown in
figure 3.
Two curves are drawn.
The upper one in the second iteration
approximation, the lower curve
is the first iteration
approximation. The spectrum
lies between these curves.


\begin{figure}[hgh]

\includegraphics[angle=270,totalheight=10cm]{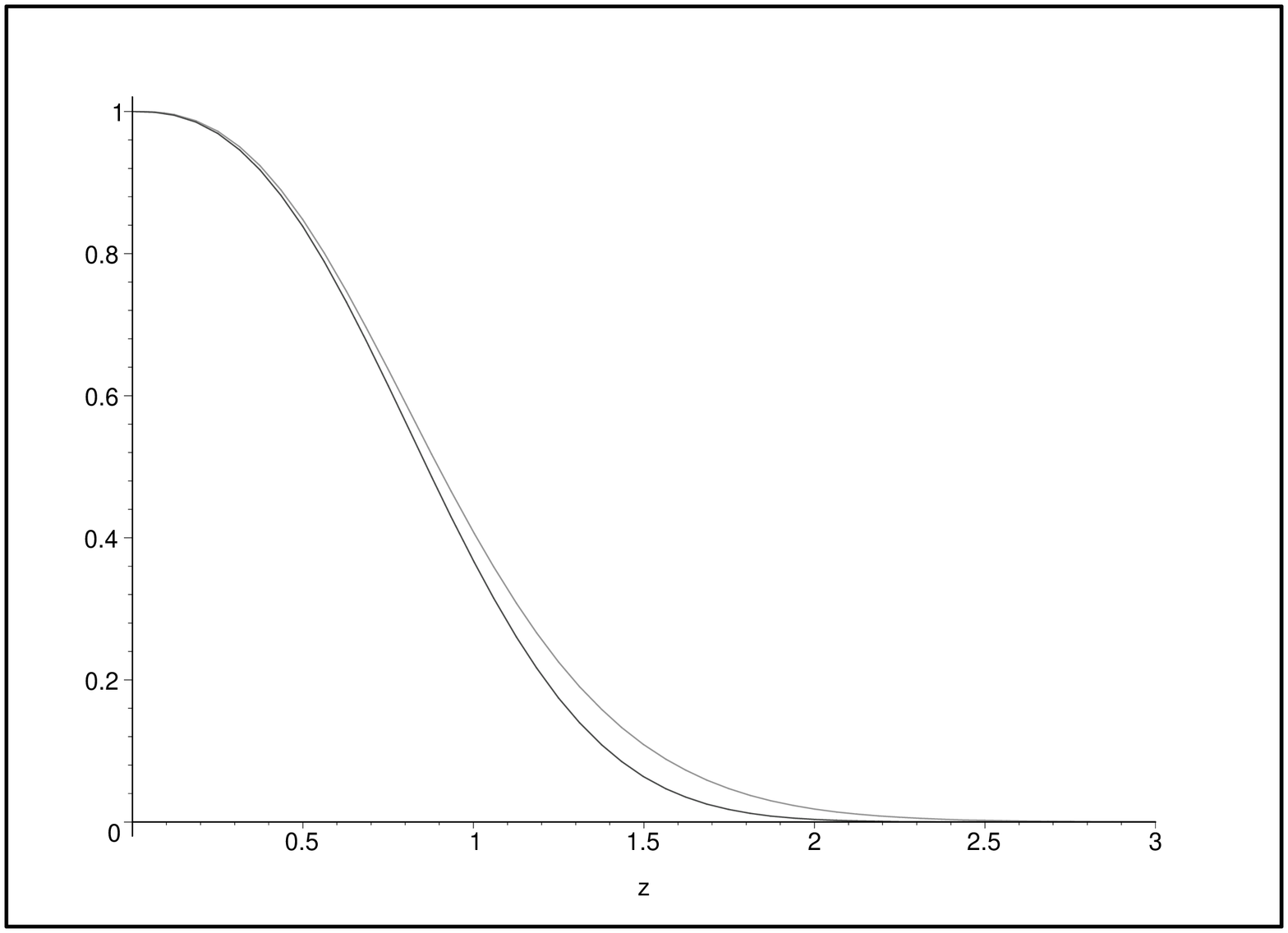}

\begin{caption}
{The form of spectrum in TAC}
\end{caption}

\end{figure}

One can see that the characteristic length
of spectrum is
$x_0 \approx 1.2 \div 1.25 $.

The total number of droplets is
$N_{TAC} = 0.9292$.
The first
iteration gives
$N_{first} = 0.8773$.

Now we investigate the two cycle model. The relative
number of droplets as a function of the boundary $p$
is shown in figure 4.

\begin{figure}[hgh]

\includegraphics[angle=270,totalheight=10cm]{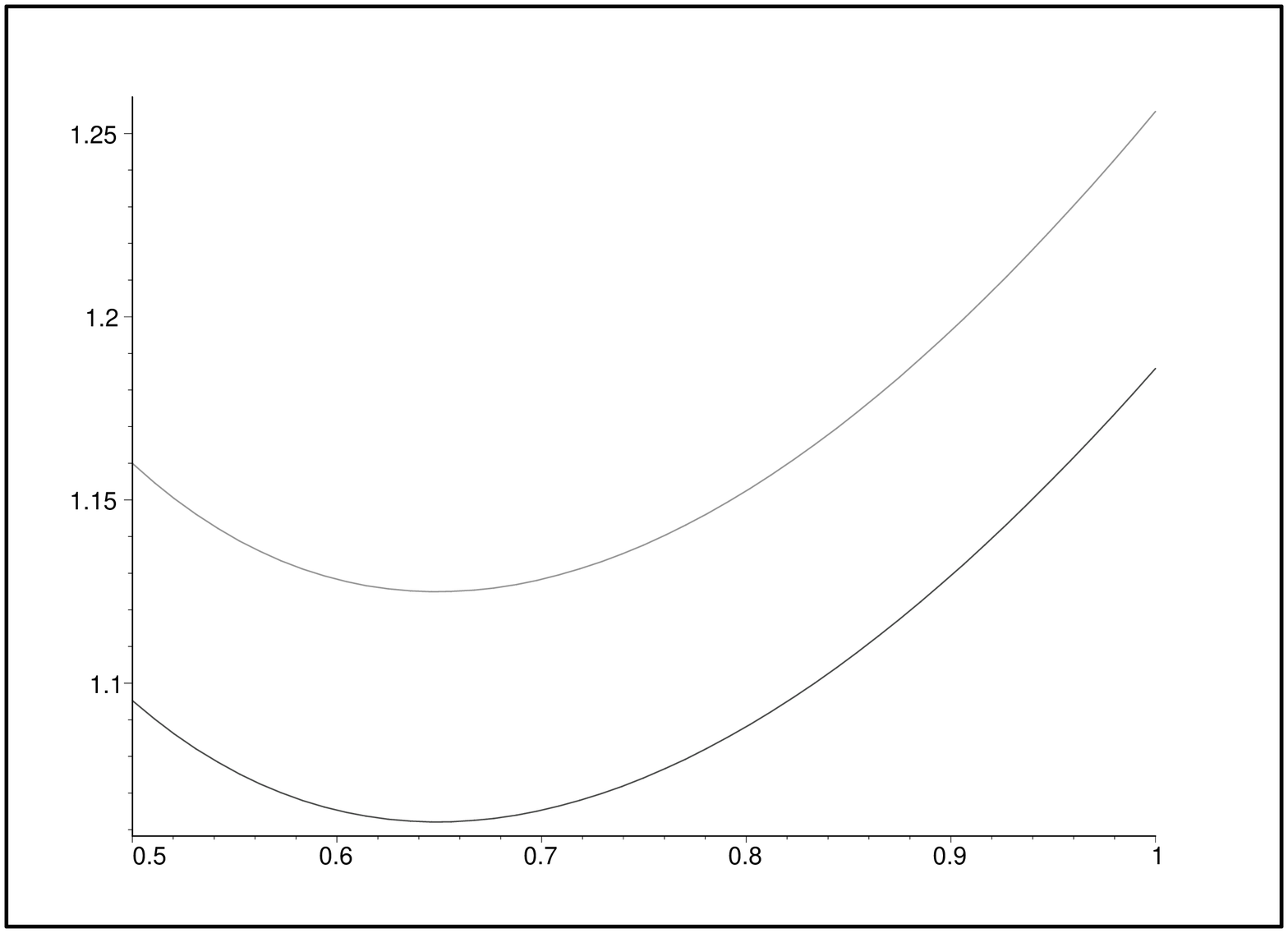}

\begin{caption}
{Mean value of the relative
droplets number in two cycle model}
\end{caption}

\end{figure}

Here two curves are drawn.
The upper corresponds to the
number of droplets referred to
the number of droplets in
the first iteration,
the lower curve is referred to the
precise value of the droplets number.
One can see that the
deviation of the second curve is two
times closer than the
deviation of the
first curve in the value of minimum.
The minimum is attained
at $p_0 =0.64$
which is approximately the $55$ percent of $x_0$
(the same part as in the free molecular regime!
\cite{statiae}).

Now we calculate $\beta$.
The values $\beta_1$ and $\beta_2$ are
drawn in figure 5.
One can see that the behavior
of $\beta$ is mainly governed
by behavior of $\beta_1$.


\begin{figure}[hgh]

\includegraphics[angle=270,totalheight=10cm]{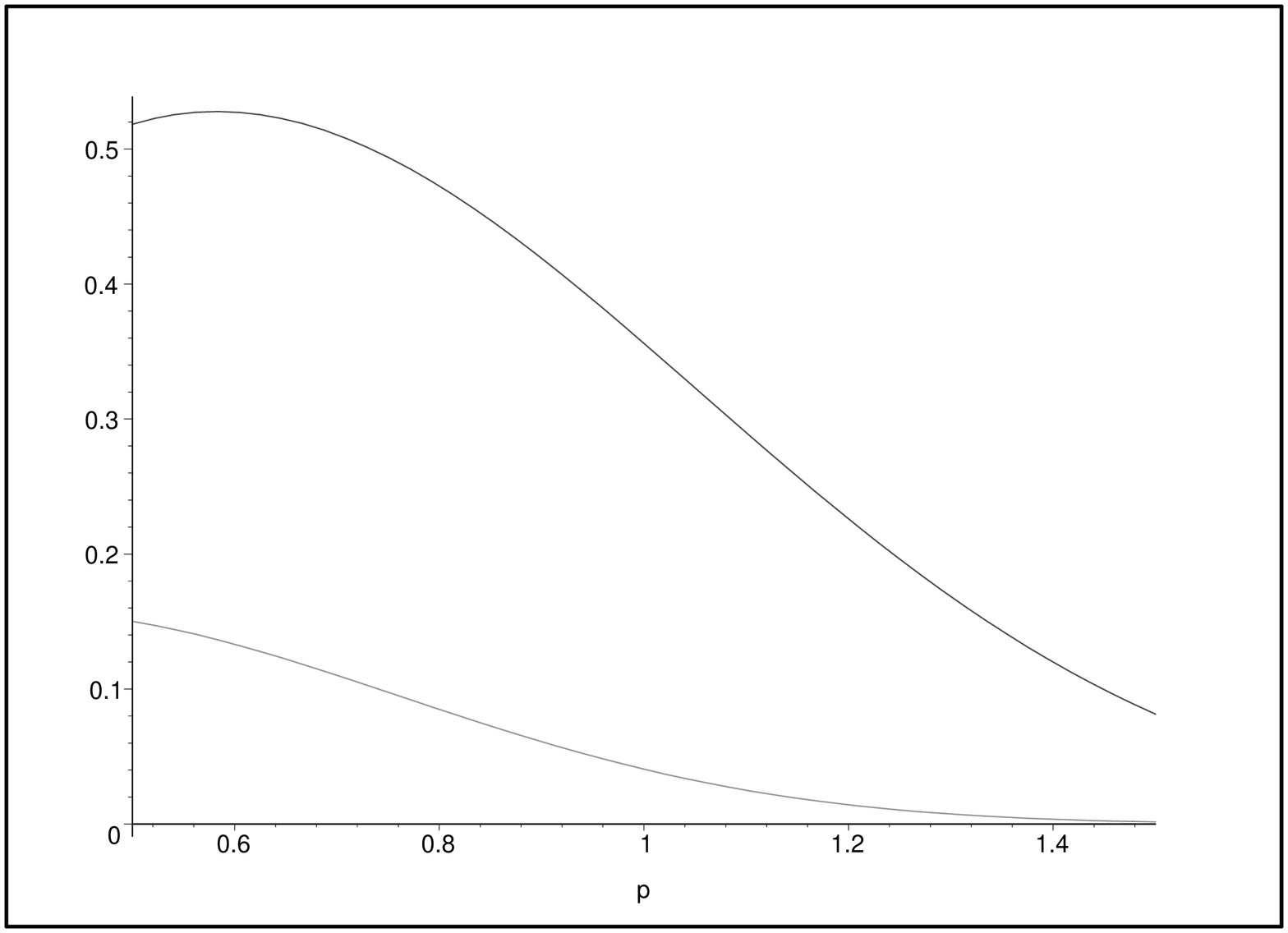}

\begin{caption}
{Values $\beta_1$ and $\beta_2$.}
\end{caption}

\end{figure}

The ratio $\beta_2/\beta_1$ is drawn in figure 6.


\begin{figure}[hgh]

\includegraphics[angle=270,totalheight=10cm]{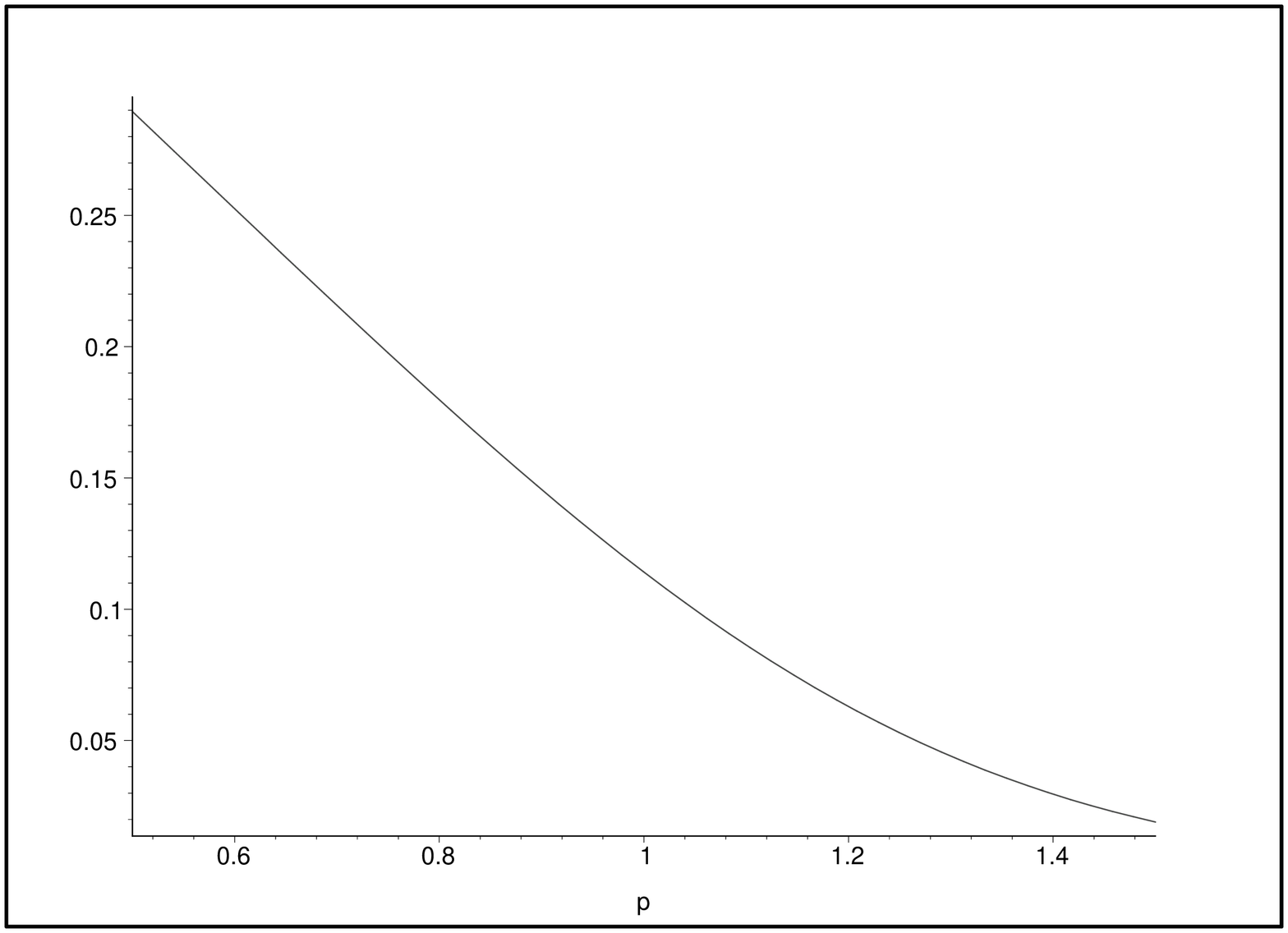}

\begin{caption}
{Ratio $\beta_1/\beta_2$}
\end{caption}

\end{figure}

One can see that there is no extremum of the ratio.
So, there is no special boundary which provides the
quickest
convergence of the chain $\beta_1$, $\beta_2$, ...
and one has to use $p_0 = 0.64$ going from
the most precise approximation for
number of
droplets as  a criterium for such boundary.
Also
one can see that the ratio is small for all $p$.

The value of relative dispersion
calculated according to
$$
D = 1 - \beta/\alpha
$$
is drawn in figure 7.


\begin{figure}[hgh]

\includegraphics[angle=270,totalheight=10cm]{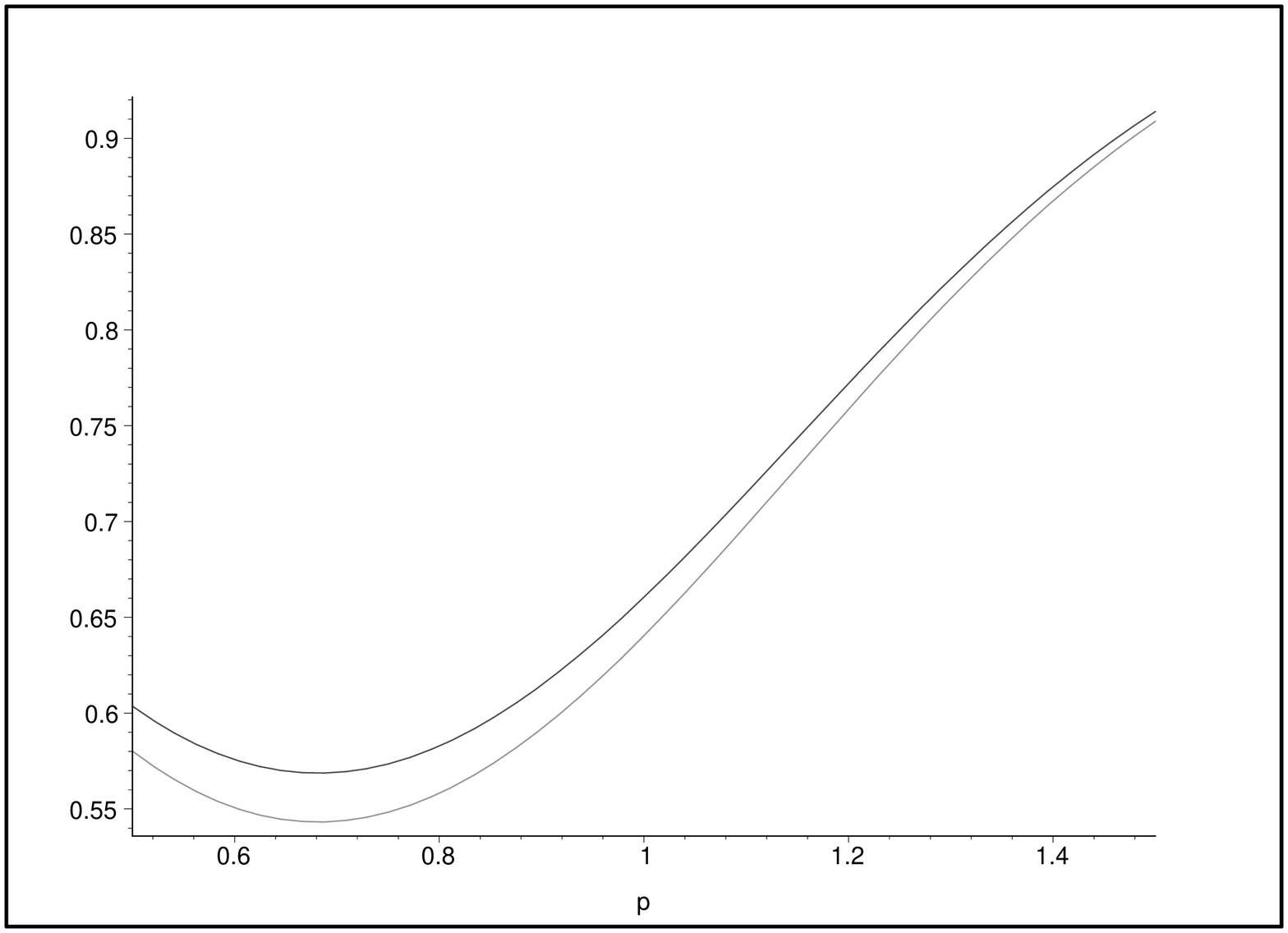}

\begin{caption}
{The relative square of
dispersion in the two cycle model.  }
\end{caption}

\end{figure}

There are two curves:
the upper is dispersion calculated
with precise number of droplets,
the lower is dispersion
calculated with the
number of droplets calculated in the
first iteration.
Both curves have extremum at $p=0.68$.
Even for the lowest curve
all values are
greater than $0.54$ while the real value
is $0.44$.

If we take into account that $\alpha$ also has to be
calculated in two cycle model.
Then the value of relative dispersion\footnote{More
precisely there is a
square of relative dispersion.} as a
function of the
boundary $p$ ini the two cycle model is
drawn in figure 8.

\begin{figure}[hgh]

\includegraphics[angle=270,totalheight=10cm]{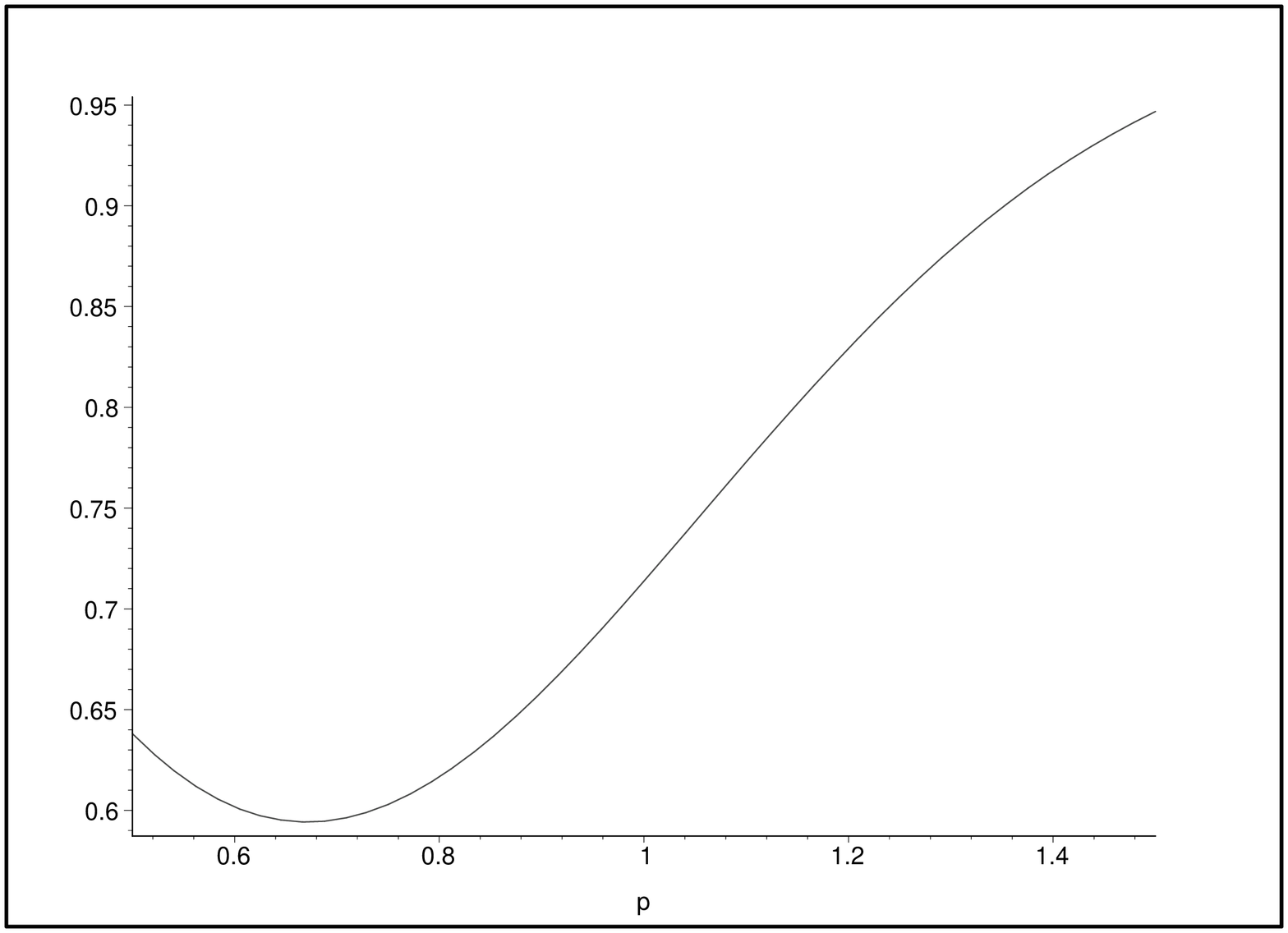}

\begin{caption}
{The relative square of dispersion in the two cycle model.
Account of the change of the droplets number.}
\end{caption}

\end{figure}

One can see that practically
nothing has been changed.
Minimum is at $0.67$,
the value of minimum is near $0.6$.

Now we try approach based on
similarity of nucleation conditions \cite{varios}.
We start with the
infinite upper boundary in formulas for
$\beta_1$ and $\beta_2$.

At first  we shall show
the result
for renormalization of dispersion when for
$\alpha$ we
take the result from the two cycle model.
The result is shown in figure 9.


\begin{figure}[hgh]

\includegraphics[angle=270,totalheight=10cm]{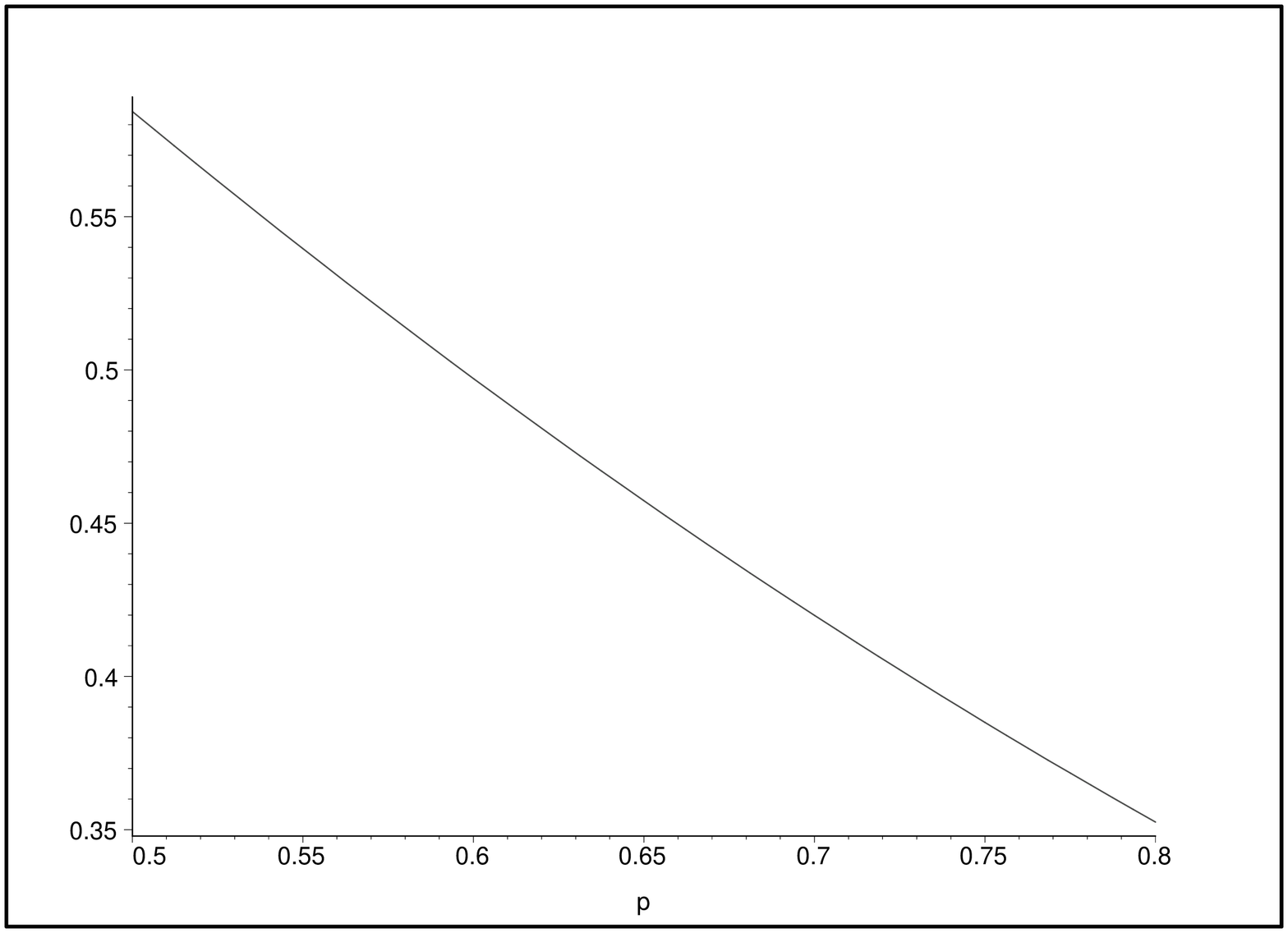}

\begin{caption}
{The relative square of
dispersion in the similarity model.
Account of the change of the droplets number in two
period model.}
\end{caption}

\end{figure}

At $p=0.64$ the value of $\psi$ will be $0.46$, at
$p=0.68$ the value of $\psi$ will be $0.43$.

When we put $\alpha = 0.92$
which is the result in the
precise solutuion we came to
the result shown in figure 10.
Practically nothing has been
changed.


\begin{figure}[hgh]

\includegraphics[angle=270,totalheight=10cm]{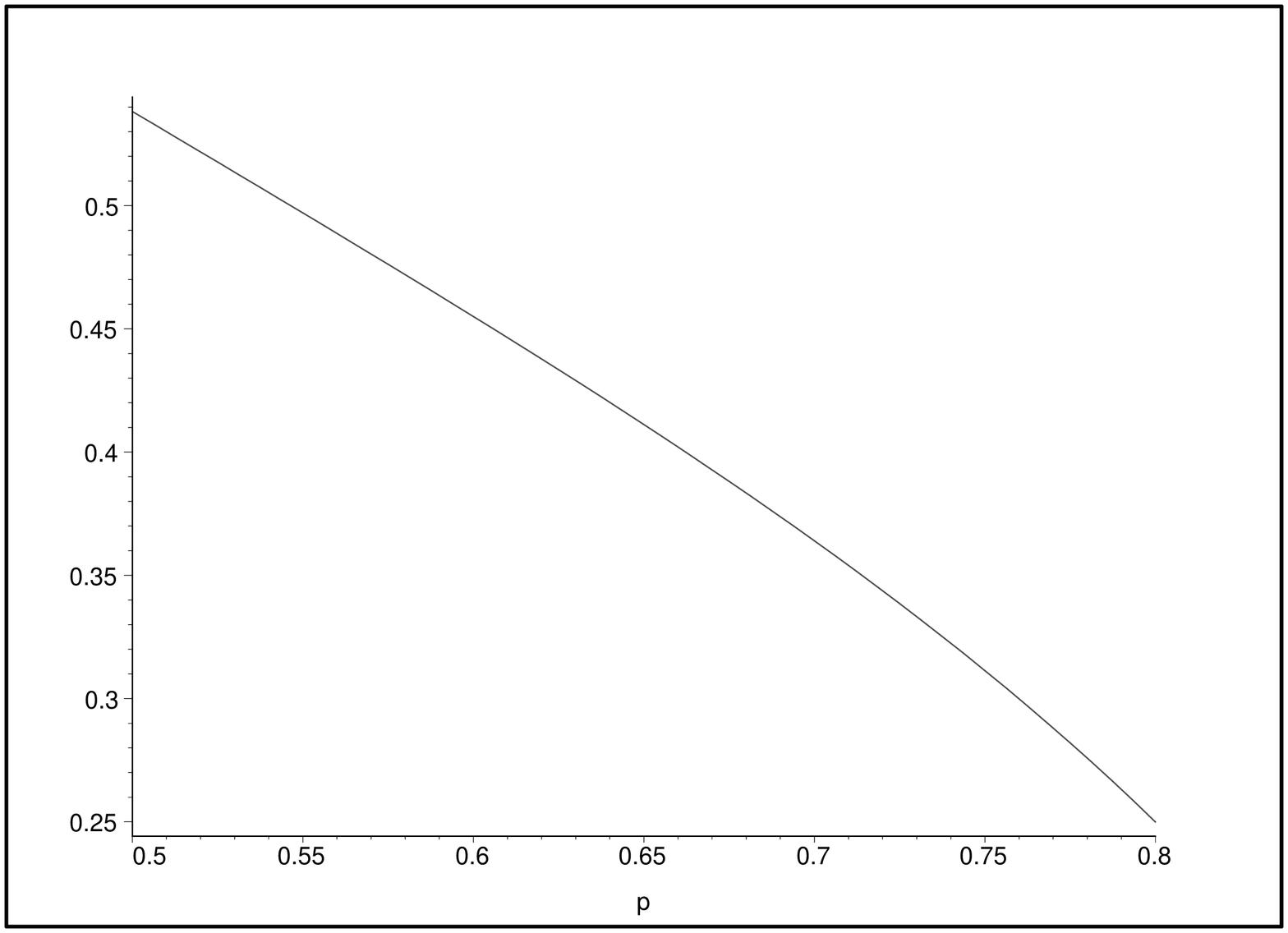}

\begin{caption}
{The relative square of
dispersion in the similarity model.
The droplets number is
taken from precise solution.}
\end{caption}

\end{figure}

Now we shall restrict the upper
limit of integration by
$z=1$ (this is the current moment of
observation). At
first we draw the number of
droplets as function of $p$
in figure 11.


\begin{figure}[hgh]

\includegraphics[angle=270,totalheight=10cm]{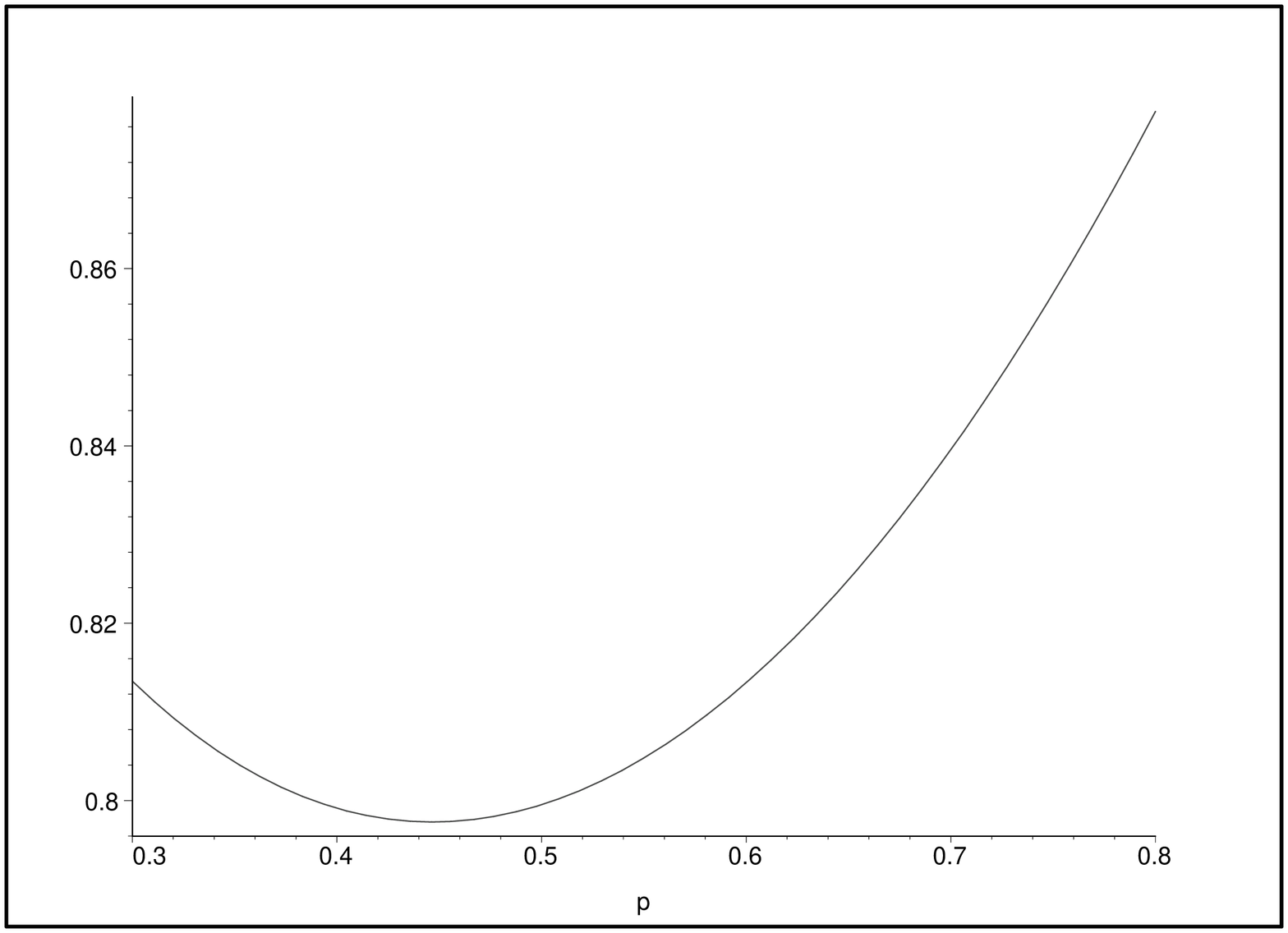}

\begin{caption}
{The number of droplets until the current moment.}
\end{caption}

\end{figure}

 We
see that the number of
droplets has minimum at $p=0.43$.
This value has to be chosen
as $p$ in calculation of
dispersion.

Then  the value of
relative square of dispersion can be
calculated and it is drawn
in  figure 12.


\begin{figure}[hgh]

\includegraphics[angle=270,totalheight=10cm]{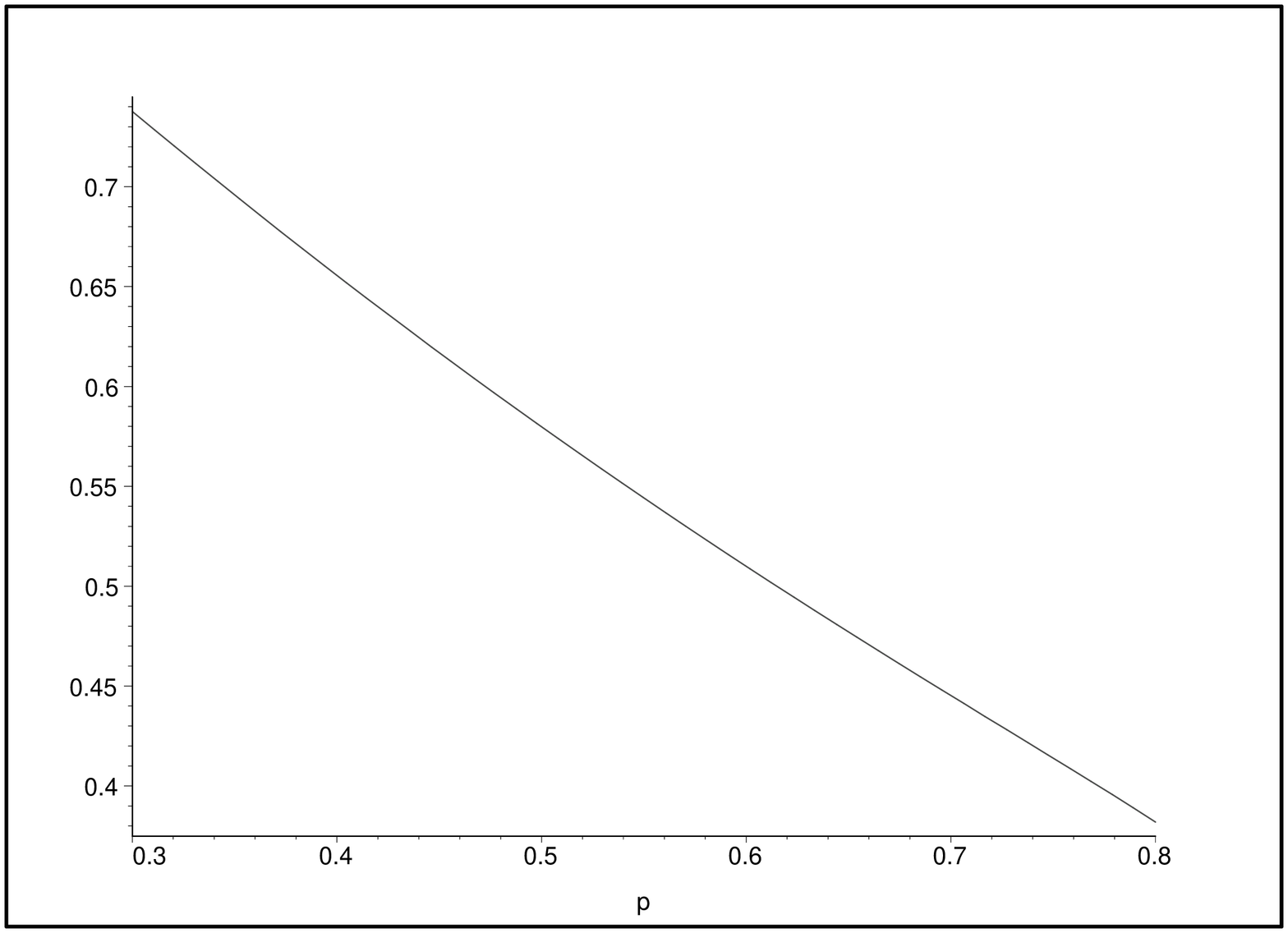}

\begin{caption}
{Relative square of dispersion up to the current moment.
Account of the change of the droplets number.}
\end{caption}

\end{figure}

We see that at $p=0.43$
the value of dispersion
is near $0.64 $ and it is far from the expected value.
It resembles the  calculation without similarity
taken into account.

It is interesting to note that
in diffusion regime the
values
of relative dispersion $\psi$
calculated in the
model with the upper boundary $\infty$
and with the upper
boundary $1$ are practically the same.
This is true for
moderate $p$ (more concrete $ p  \leq 0.7$).

An explanation of this
property is the following: under the
free molecular regime of
droplets growth the droplets
appeared in the first moments
of time grow so rapidly that
it was very
important  to know until what sizes (or until
what moment)
they will grow. So, the duration of the
nucleation period
was very important.  Under the diffusion
 regime of growth
 this effect is much more weak. So, the
 position of the boundary isn't so essential.

For $p \leq 0.5$
the value $1-\beta / \alpha$ is also close
to these values.

The explanation
lies also in collective character of
stochastic effects
initiated under the diffusion regime of
droplets growth.

The question to decide is
whether we have to take $p$ from
extremum
for $\alpha$ calculated with the upper boundary
$\infty$ or with
the upper boundary $1$. From the first view
it seems that
the boundary $1$ is preferable because
$\beta_1$ and $\beta_2$ are
calculated with the upper
boundary $1$.
But the extremum for $\alpha$
corresponds to the
value $\alpha = 0.815$. This value is
smaller than $\alpha$
calculated in the first iteration,
i.e. $\alpha_1 = 0.8773$ and
smaller than this value
calculated in the
precise solution $\alpha_{pr} = 0.9292$.
So, at extremum
(it is  a minimum) of the model with the upper
boundary $1$
the deviation from the precise value
attains maximum.
At the upper boundary $\infty$ the value
of $\alpha$ is always bigger than $1$ which is
greater than
$\alpha_1$ and
$\alpha_{pr}$. So, here
the extremum is really the closest
value to precise solution.
Then it has a real physical meaning.
So, it is reasonable to take $p$
corresponding to the
extremum of $\alpha$ with the upper
boundary $\infty$.

Calculations give $p = 0.65$
and $\psi = 0.49$. This value
is calculated
on the base of $\beta_1$ and $\beta_2$ with
the
upper boundary $1$.
The value $\psi = 0.49$  is rather close
to result of numerical
simulation $\psi = 0.44$.
Certainly, the relative
error is one tenth and the methods
based on the
explicit two cycle models cannot give more precise
results.

One can note the following curious result.
Evidently, when we calculate
$\psi$ for $p= 0.65$ with the
upper boundary $1$
we come to $\psi = 0.46$ which is
practically the necessary result.
But this coincidence is
no more than an
occasional one. When we try the same
procedure for the
free molecular regime we come to $\psi=
0.55$ (here $\psi=0.55$)
which is far from numerical
result $\psi = 0.64$.

Having estimated the value of
deviation of $\psi = 0.49$ from
$\psi = 0.44$ one has to note
that it is much smaller than
the deviation of $\psi = 0.59$
of the value $1 - \beta / \alpha$
in its minimum from $\psi = 0.44$.
So, the modification made
here really have a real sense.

\section{The effect of
growing volumes of interaction}

What shall we do to
take into account the growth
of the volume interaction
in diffusion regime of
growth?
It is necessary  to rewrite
formula
(\ref{starting}).
We can
substitute in
$ \exp(-\frac{(N_i -
\bar{N}_1)^2}{2 \bar{N}_1} ) $ the
denominator $2 \bar{N}_1$ by another denominator
which corresponds to the growing volume of
interaction. Then $\bar{N}_1$ has to be
substituted by $\bar{N}_1 const (j-i)^{5/2}$ with
some
known $const$. It means that the characteristic
half-width $\sqrt{\bar{N}_1}$ has to be
substituted by $\sqrt{\bar{N}_1 const (j-i)^{5/2}
}$. One can not apply directly this substitution
in
$ \exp(-\frac{(N_i -
\bar{N}_1)^2}{2 \bar{N}_1} ) $
but only
approximately
in  expression for $a_i^{(k)}$ which
now
will be
$$
a_i^{(k)} =
1- \frac{ 5/2 }
{ M^{5/2} }
\sum_{j=P+1}^k
\exp(-\frac{j^{5/2}}{M^{5/2}})
(j-i)^{(\frac{1}{1-s} +1) - 5/2}/{const}
$$
with some
known $const$ and an appropriate renormalization.

In diffusion regime
$$
\frac{1}{1-s} +1  - \frac{5}{2} = 0
$$
and there is absolutely no effect.
It means that kinetics
of nucleation will be
absolutely different. It is described
in \cite{PhysicaA}.
The most strong effect will take place
in the free molecular regime where
$\frac{1}{1-s} +1  - {5}/{2} = 3/2$

Now we shall estimate the effect of the
dispersion variation.
We see
\begin{eqnarray}
D^{(\infty)} \sim
\tilde{N}^{(\infty)} (1-\frac{1}{\alpha}(
2 (\frac{1}{1-s} +1)
\int_0^{p} dx \int_{p}^{\infty}
dz (z-x)^{\frac{1}{1-s} +1 - 5/2} \exp(-z^{5/2})
\nonumber
\\
\nonumber
- (\frac{1}{1-s} +1 )^2
\int_0^{p} dx (\int_{p}^{\infty} dz
(z-x)^{\frac{1}{1-s} + 1 - 5/2}
\exp(-z^{5/2}) )^2 ))
\end{eqnarray}
Note that the last relation has to applied to the
subsystem with the size $\sqrt{4Dt_n}$ where $t_n$
is the duration of the nucleation period. Then we
have to take the superposition of these independent
subsystems.

We see that
in the majority situations
we can approximately neglect
$(j-i)^{\frac{1}{1-s} +1 - 5/2}
\approx (j-i)^0 =1$ and
see that then
$$
D^{(\infty)} \approx
\tilde{N}^{(\infty)}
$$
and there is no effect.

Nevertheless the deviation in the number of
droplets from the
value calculated on a
base of averaged characteristics
 exists. We have to divide the whole
system into subsystems of the size
$\sqrt{4Dt_n}$, then
determine the  number of droplets in this subsystem
$\tilde{N}^{subs}_{tot}$ obtained on the averaged
characteristics, then determine
the deviation
$\delta{N}^{subs}_{tot}$
from the values calculated in TAC and then
the whole deviation
in the whole number of
droplets $N_{tot}$ from the
value based
on the averaged value
$\tilde{N}_{tot}$ will be
$$
\delta{N}_{tot} =
\delta{N}^{subs}_{tot}
\frac{\tilde{N}_{tot}}{\tilde{N}_{tot}^{subs}}
$$

These procedures completely solve the problem.

One has also to note that the use of monodisperce
approximation \cite{Monodec} leads to the absence
of the
problem of growing
volume because the characteristics
governing the nucleation period
have been determined only
in one moment of time.
The problem here is a justification of
monodisperse approximation to the
calculation of fluctuation effects.

\section{Appendix}

Here we shall present
the
two cycle model with a fixed boundary in
kinetics of the metastable phase decay
under the free molecular
regime of the droplets
growth. This question
was completely
investigated in
\cite{statiae} but
during the last two
years our view on this
problem was slightly
modified and the modern
state is described below.

Systematic investigations of the first order phase
transitions are performed since Wilson \cite{Wils}.
  The classical theory of nucleation
\cite{class} gave expressions for all
main characteristics of stationary
process of nucleation.
This allowed to investigate a global picture
of the phase transition. A set of papers
\cite{Kinet} was devoted to  model pictures of the
global kinetics of nucleation.
Here we shall also consider the global
picture of the phase transition.
One can note that  all
cited publications \cite{Kinet} were
based on averaged nucleation rates. Here we shall
consider stochastic appearance of embryos and outline the
 stochastic manner of appearance.

Recall briefly the main features of  phase
transition. Suppose that in initial moment of
time there exists a metastable state. Then
 the embryos of a new phase
begin to appear in the
metastable system. The average rate of appearance
is given
by \cite{class}. Then the embryos begin to grow
and to consume the vapor, metastability falls
and the rate of nucleation, i.e. the rate of
appearance of new embryos falls also.
The vapor consumption occurs in a time scale
in a
very sharp avalanche manner.

 It is clear that during the nucleation period
 the new  supercritical
formations of the new phase appear
with some fixed
probability, but they appear in stochastic
manner.
So, the stochastic
appearance of relatively big number of droplets
leads to very rapid stochastic consumption of vapor.
Stochastic
appearance
of relatively small number of droplets leads to
delay of the vapor exhaustion
and to excess of
droplets appearance in next moments of
time.
But
it seems that this excess  can not compensate the
opposite effect of the absence of embryos in
first moments of time.
It seems that the
total number of droplets will differ from the
average value.
This is the naive reason why stochastic
effects of nucleation have to be taken into
account.

In a system with macroscopic sizes  due to a giant
value of the Avogadro
number there appears some rather
big number of droplets.
It allows to use the averaged characteristics to
construct  kinetics of a nucleation process.
Precise kinetic approach
based on averaged  characteristics
is described in \cite{PhysRevE94}.
In \cite{PhysRevE94} the time evolution is
completely constructed.

After the formulation of
integral equations (see \cite{PhysRevE94})
one can introduce
"elementary intervals of nucleation" -
the intervals where the
state of the system changes negligibly small.
In macroscopic systems the
total number of droplets is so big that
at every elementary interval there appears a great
number of droplets   $\Delta N$.
On the base of traditional thermodynamics
one can state that the relative fluctuation
 $\delta \Delta N/ \Delta N$
of droplets formed at elementary interval is
small and has an order of
$(\Delta N)^{-1/2}$.
This remark completely solves a problem of
justification of nucleation description based on
averaged characteristics.

Since
it is possible to extract elementary intervals
where
thermodynamic  parameters and the nucleation rate
have small variations
there is no need to take care about
the stochastic corrections.

In experimental investigations one can not
study  quite
macroscopic systems because the most popular
data is the
number of droplets and to fulfill the
calculation
of this number one can
not have too many
droplets\footnote{One can not simply
calculate the infinite
number of droplets. The upper
limit of the number of droplets
which can be noticed on a photo image is about
several dozens.}.
After the theory based on the averaged
characteristics
has been presented it became possible to
investigate the
stochastic effects in kinetics of nucleation,
i.e. the
effects of stochastic appearance of droplets.
Recently there appears a set of papers \cite{Vest},
\cite{Koll}, \cite{Kolldyn}
where a stochastic effects (the effects of
fluctuations of droplets formation) were
described and investigated.
One can extract two approaches which were
formulated in \cite{Vest} and in
\cite{Koll}. Although \cite{Koll},
\cite{Kolldyn} were written
by the same authors of \cite{Vest}
these authors didn't hesitate
that the theories formulated in
\cite{Koll} , \cite{Vest}
gave different results.
So,
at least one has to analyze approaches
\cite{Vest}, \cite{Koll}
and decide whether there is a true
result among these
approaches
and how it can be used to construct the
adequate description of the nucleation process.

One has to specify a formulation of the
problem of our
investigation.

In \cite{Koll},  \cite{Vest}
it was proposed to establish
corrections to the total
number of droplets $N$
appeared in the system.
It was supposed that these corrections  are
functions of
$N$.
To demonstrate the error of this approach it is
sufficient
on one hand
to take two identical systems then to
calculate them separately and to add results or
on the other hand
to calculate correction directly for the total
system. The results will be  different.

One has to determine a real volume to which one
has to refer the number of droplets.
It is simple to do with the help of results from
\cite{PhysicaA}.
In that paper  the kinetics of nucleation for
diffusion regime of droplets growth was
constructed. It was shown that a solitary droplet
perturbs vapor up to distances of an order
$\sqrt{4Dt}$,
where $D$
 is  a diffusion coefficient,
 $t$
 is a time from
 a moment of droplet formation up to a current moment.
 As an estimate one can take as
    $t$
 a time
  $t_1$
  of the whole nucleation
  period duration\footnote{The nucleation
  period is a period of intensive formation of
  droplets.}.

The last remark
allows to give a new definition of the volume
 $V_{el}$
where the number
$N_{el}$ of interacting (mainly through
the vapor exhaustion) droplets is formed.
Namely this value has to be regarded as a volume
of a system in approach \cite{Vest}, \cite{Koll}.
This volume is
$$
V_{el} = 4\pi (4D t_1)^{3/2}/3
$$
If the sizes of the system are
smaller than this value
one has to take the volume of the system as this
value.
But such a small system can be hardly
regarded as
a macroscopic one. At least one has to analyze
carefully the boundary conditions for the system.

Naturally, the droplets appeared at different
moments of time perturb initial phase up to  different
distances. So, one can regard above formulas only
as estimates. Some more rigorous equations can be
found in
\cite{book2}.

The number of droplets  $N_{el}$
isn't too big as
$N$ is.
So, an analysis of stochastic effects has a real
sense. It is interesting now to get all
correction terms which are ascending with the
number of droplets (but not  only a leading
term).
To solve this task one has to modify approaches from
\cite{Koll}, \cite{Vest}.

Complexity of this problem appears here because
 one can not directly use equations based on
the theory with averaged characteristics. In
\cite{Koll},
\cite{Vest}
some properties of solution of the
theory
on the base of
averaged characteristics (TAC) were the
starting points for
constructions.
This supposition was adopted without any
justification.
So, at first we have to decide whether it is
possible to
start with TAC.

We shall consider the situation of decay
 of metastable phase. The new
dimensionless
parameter - the number of droplets destroys the
universality observed in
\cite{PhysRevE94}
for the theory based on averaged characteristics.
Moreover, it is difficult even to formulate the
system of equations. This  radically complicates the
problem.

The possibility to use
the effective monodisperse approximation
formulated in
\cite{Monodec}
was used in  \cite{Vest}
without any justifications. Generally speaking
this property can not be directly used to
calculate
stochastic corrections.
One has to analyze
whether this conclusion leads to an
error made in
 \cite{Vest}.
 Here we have to formulate more correct
 constructions.

Both approaches from \cite{Koll} and from
\cite{Vest} declared that they
used the following property
observed in \cite{PhysRevE94} in frames of TAC:

{\it "The droplets formed at the beginning
(i.e. at the first half) of the
nucleation period are the main consumers of
vapor". }

This property is valid \cite{PhysRevE94},
but it is substituted in reality in
 \cite{Vest}, \cite{Koll}
by the following statement:

{ \it The main source of stochastic effects are the
free fluctuations of
droplets formed at the beginning
(i.e. at the first half) of the
nucleation period. They govern the fluctuations
of all other droplets}.

The last statement
seriously differs from the first one.
To get credible
results it was necessary to balance the
fluctuation
effects from the first half of spectrum  by
the corresponding reaction of remaining part of spectrum.
Then,
at least
this approach
needs some  special justifications.
So, one has to use some new constructions which are
presented below.

The
application of the model
approximation
which was in reality done in
\cite{Koll}, \cite{Kolldyn}
will lead to some errors. But due to
universality of solution \cite{PhysRevE94}
the errors  can not be be too big.  Qualitatively
everything is suitable, but precision
will  be  not so high.

The same conclusion will be valid
for
any approach  based on some model
behavior of supersaturation (justification is
valid for a vapor consumption in TAC, but not for
stochastic effects).
Namely, in \cite{Koll}
was used an
artificial approximation where at the first
half of the nucleation period
the conditions of nucleation
are the ideal ones and at the
second half the conditions of
nucleation are governed by the
droplets formed at the first
half.
Here in current paper the final
result will be more precise and it
will be not based on  rather
spontaneous artificial choice
of some parameter as it was done in \cite{Koll}
where this parameter  was put to
to $1/2$.
In \cite{Koll}
it is supposed that until some moment
(it is chosen in  \cite{Koll}
as a half of nucleation period) the droplets
are formed under ideal conditions and namely
these droplets determine a vapor consumption.
In reality this approach taken from
\cite{book1} (page 310)
was used in  \cite{Koll}
in slightly  another sense.
It is supposed that droplets formed during a
 first half of nucleation period
 are the main source of stochastic effects.
The last statement was not justified in  \cite{Koll}
and it is rather approximate. The relative
correctness  of a result was attained due to
specific compensation of different errors of
approximations used in
\cite{Koll}.
It is necessary to stress that the the
mentioned model was
used in \cite{book1} to justify a strong
inequality and the high
precision of constructions was not
essential.
But in \cite{Koll} this model was the source of
numerical result.

All arguments listed above lead to necessity of
reconsideration which will be made in this
publication.
A plan of narration will be the following one
\begin{itemize}
\item
Having considered the interaction of stochastic
deviations of the  number of droplets appeared
during the elementary
intervals of nucleation we shall see
that stochastic effects are
at least moderate.
\item
A
moderate scale of stochastic effects allows to seek
the solution on the base of the theory with
averaged characteristics. But we have to take
stochastic effects from all droplets formed
during the nucleation period.
\item
The possibility to take into account the influence
of stochastic deviations of
all droplets can be provided by
the property of the self
similarity of nucleation conditions during  the
nucleation period. This property can be
considered in two senses - 1) in the local
differential sense  and 2) in the integral sense
in frames of the first iteration in the iteration
procedure in TAC \cite{Novos}. The local
property will be used in justification of the
smallness of stochastic effects and the integral
property will be used to calculate the
stochastic corrections of the whole nucleation period.

\end{itemize}

All analytical results will be checked by
computer simulation and a coincidence  will be shown.

All mentioned constructions will be valid for an
arbitrary first order phase transition.
The law of droplets growth here will be a free
molecular one, then the linear size
of droplet grows with
velocity independent from its value.
Consideration of other regimes can be
formally attained in
frames of the current approach by some trivial
substitutions, but one has to take into account
that the new regime requires new approaches to
construct nucleation kinetics as it is shown in
 \cite{PhysicaA}.
So, one can not agree with  \cite{Vest}
where it is stated
that one of results is an account of stochastic
effects in a diffusion regime of droplets growth.
This effect has to be
taken into account principally in
another
manner by
application of methods presented in
\cite{PhysRevE2001}.

\subsection{Some characteristic
features of decay kinetics}

We begin with  the theory based
on averaged characteristics.
It is supposed to be known \cite{PhysRevE94},
that the supersaturation $\zeta$
behavior can be determined after
certain
renormalizations by the following equation
$$
\psi (z) = \int_0^z dx (z-x)^3
\exp(-\psi)
$$
on function
$\psi$ which is the relative renormalized
deviation
of supersaturation from the initial value.
Variables
$x$ and $z$ can be considered as equivalent ones.
A good approximation for solution and for
a distribution  of the droplets number
over time (or over liner sizes of droplets)
which is proportional to $\exp(\psi)$
 is
\begin{equation}
 \label{1}
f_1 =
\exp(-z^4/4).
\end{equation}
The form of $f_1$
is given by fig.13.
\begin{figure}[hgh]

\includegraphics[angle=270,totalheight=10cm]{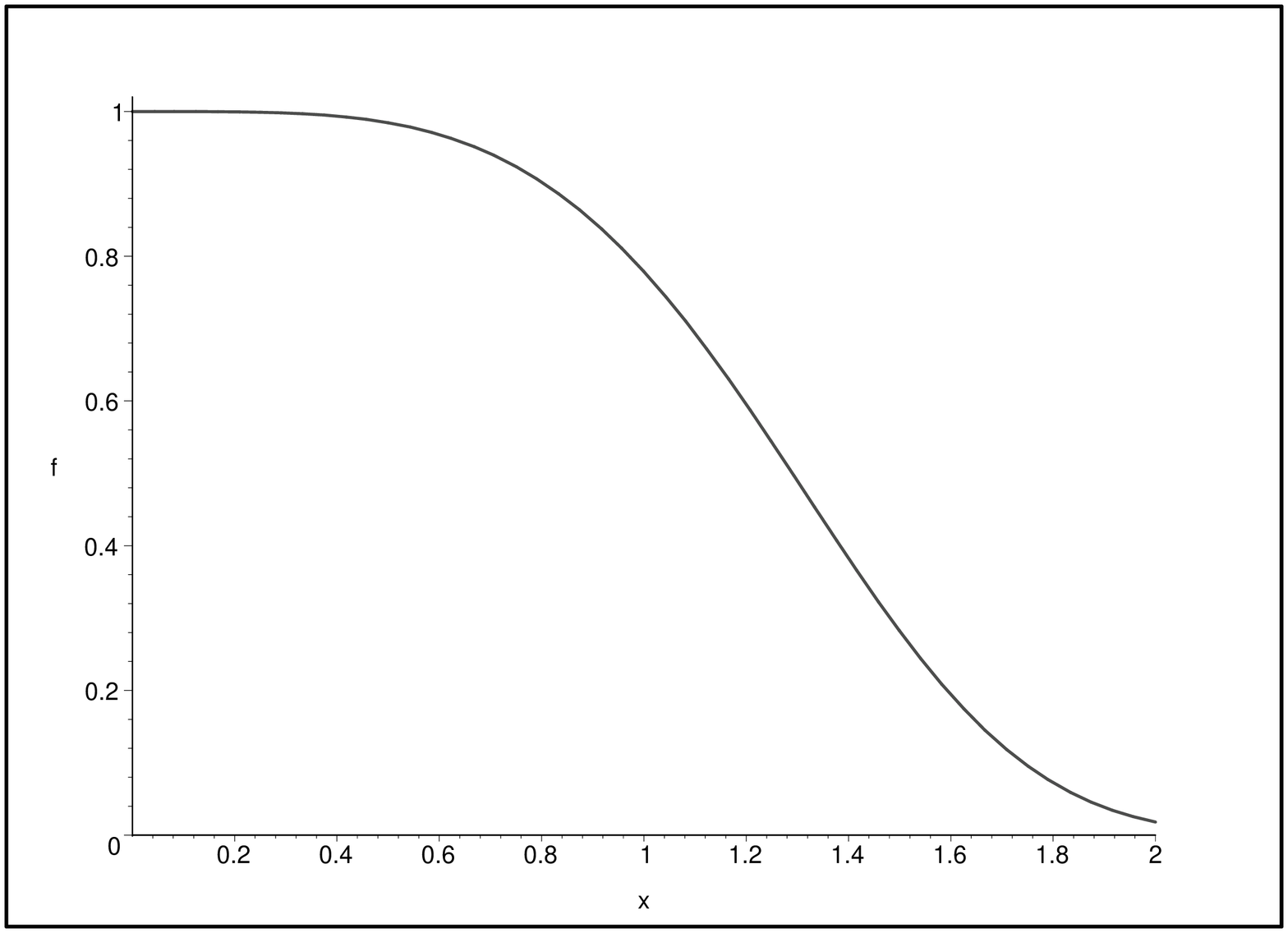}

\begin{caption}
{
A form of approximation for size spectrum}
\end{caption}

\end{figure}
It is seen that approximately  at $z_0=1.25$
the nucleation period stops.

Approximation (\ref{1}) has rather high precision
\cite{Monodec}.
 It is based on the following law of substance
accumulation
$$
G =
z^4/4 = \int_0^z (z-x)^3 dx \equiv \int_0^z g(x) dx , \
\ g = (z-x)^3 \equiv \rho^3
$$
for the renormalized number $G$ of molecules in a new
phase.
For any moment $t$
or $z$
a function $g$
has one and the same form.
We shall call this property as  a
"similarity of nucleation
conditions". We see that every time the droplets
formed at the last third of a period from
beginning of nucleation until a current moment
will accumulate a negligible quantity of
substance.
The relative quantity of the substance there has
an order of
($ \sim 1/27$)
and is so small that even if there will be
fluctuations the quantity will remain small.

From the form  of $f_1$
it is seen that until
$z_= \equiv 0.7 z_0$
all droplets  deplete vapor rather weak. It
will be important for future analysis.

Another important
property is the possibility to describe
kinetics in
frames of TAC with the help of monodisperce
approximation (see \cite{Monodec}).
The mentioned property of $g$
allows to use a monodisperse approximation
\cite{Monodec}
not only at the end of nucleation but in every
moment of the nucleation period
\cite{Monodec}.
Let $t(G)$
be the moment when
there are $G$ molecules in droplets
(in appropriate units).
An application of the monodisperse approximation
 \cite{Monodec}
leads to
$$
G \approx  N_m(z) z^3
$$
where $N_m(z)$ is
the number of droplets born until $z/4$
$$
N_m (z) = N(z/4)  \approx z/4
$$
and
$$
N_m(z) \approx \frac{N(z)}{4}
$$
for $z$ which are essentially
less than those corresponding
the end of the nucleation period (in reality
$z \leq z_{=}$.

\subsection{Interaction of arbitrary stochastic
fluctuations}

The account of fluctuation
interactions at every moment of
time is important in
justification of the smallness of
stochastic corrections.
Generally speaking, one can not
take consideration only at
the end of nucleation period
without justifications that
the characteristic features of
nucleation kinetics remains
the same at every moment of time.

The arbitrary value of $z$ in
the period of nucleation
corresponds to the
arbitrary value of the number of
droplets in a
liquid phase $G$ (in renormalized units
which will be
used (see \cite{PhysRevE94}) the value  $G$
will belong to interval $(0,1)$).
Now it will be possible to repeat in some features
the approach from
   \cite{Vest}
with arbitrary parameter $G$ instead of $1$
(in renormalized units, before renormalization
it would be $1/\Gamma$  (see \cite{PhysRevE94})).
The sense of a difference from  \cite{Vest}
is consideration of an arbitrary moment $z$ instead
of the end of nucleation. It is very important because
allows to take into account all fluctuation effects
during the nucleation period.

Let $t(z)$ be the current moment of time
  ($z$ is the coordinate of the spectrum front,
actually $t$ is proportional  $z$).
We suppose that before $az$ ($a$
 is some parameter)
droplets are formed without mutual influence and
one can write Poisson's distribution.
This is  the first group of droplets.
The second group of
droplets are all other droplets formed until the
time
moment $z$.
A natural restriction on
$a$
appeared, namely $a< 0.7$.
We
shall take $ a > 1-0.7 = 0.3$ also
for the purposes explained below.
We suppose that the influence of other droplets
on its own formation is negligible
(this follows from
$2 * 0.7 = 1.4 > 1.25$
and from notation made above about the last third of
nucleation period).
Then it is possible to write Poisson's
distribution for the second group of droplets,
but with parameters depended on stochastic
values - characteristics of the droplets
distribution from the first group. Rigorously
speaking one has to use
the first four moments of the droplets distribution
in accordance with \cite{PhysRevE94}, but for
simplicity we shall use here only the zero
momentum. As a compensation for this simplicity
we has to use here only
$a=1/4$
which corresponds to the applicability of monodisperse
approximation. But due to the arbitrary value of $G$
this is quite sufficient for our goal.

Certainly, one can not state that
precisely the first fourth
of the spectrum plays the main role
in vapor consumption.
So, we have to consider interactions of stochastic
fluctuations for all $a$ which aren't too small in
comparison with $1/4$ as well as $1-a$ isn't
too small in
comparison with the same $1/4$.
This will be done below.

At the next step of consideration
one has to come from
Poisson distributions to Gauss distributions
and integrate them with account of connection
between stochastic parameters of embryos
formation from the first group and parameters of
distribution from the second group.
The same was done
in \cite{Vest} but only for a leading
term.

Why it is necessary to get all ascending
terms in
corrections? The existence of at least one
coefficient with a
big absolute value
means (see the theory of Chebyshev
polynomials) that there exist a
size of a group where the
interaction will lead to the
big effect. Then it would
lead that the interaction in these
groups will be the real
driving force of the process
and these groups can be
regarded as quasiparticles.

Contrary to \cite{Vest}
we shall take into account all correction terms
which come from transition from Poisson's
distributions to Gauss
distributions and corrections for
nonlinear connection between the group distributions.
We shall take all terms which are growing when
the total number of droplets grows.

We  get the following result for droplets
distribution
$$
P = P_{\infty}
(1+ y)
$$
where
$$
P_{\infty} = (\frac{9 a}{2 \pi N (3a+1)})^{1/2}
\exp(-\frac{9a}{3a+1} \frac{D_s^2}{2})
$$
$$
D_s = \frac{\hat{N}-N}{\sqrt{N}}
$$
$\hat{N}$ -
 some stochastic  value of the total number of
 droplets,
$N$
- the mean value of droplets and $y$ is the
 correction for spectrum.

At $ à = 1/4$
we get
$$
y=
\frac{1}{74088} D_s (8087 D_s^2 - 10269) s
+ (-\frac{4}{9} + \frac{305}{1176} D_s^2
$$
$$
 -
\frac{85903}{12446784} D_s^4 +
\frac{65399569}{10978063488} D_s^6 ) s^2
$$
where
$$
s \equiv 1 / \sqrt{N}
$$
is a small parameter of decomposition.
To
get all ascending (with $N$)
corrections we must fulfill
decomposition until
$s^2$.

Why it was necessary to get all
ascending terms in decomposition?
The answer lies in specific sequential influence
which can be observed in nucleation period. The
droplet appeared in the first moment of time
forms condition for the embryos appearance in the
second moment, then the embryos appeared in the
first and in the second moments form conditions
for further appearance, etc. So, if there would be
some $N$ at which corrections will be big then
immediately  namely this value will be the
crucial value for all kinetics. Fortunately here
there is no such effect and, thus, we can take
the theory with the averaged characteristics as
the base for further constructions.

At arbitrary  $a$
we get for $P/P_{\infty}$ (here $P_{\infty}$
is the limit at $N=\infty$)
the following expression
$$
P/P_{\infty} = 1+ w_1 s+ w_2 s^2
$$
Here
$$
w_1 =
-\frac{1}{6} (( 486 l^{12} + 486 D_s^2 l^{12} -
972 l^{11} D_s^2 + 324 l^{10} - 648 D_s^2 l^{10}
+ 756 l^9
$$
$$
+ 810 l^9 D_s^2 - 459 l^8 + 297 D_s^2 l^8 -
135 l^7  D_s^2 -
$$
$$
387 l^7 + 90 l^6 D_s^2 - 153 l^6
+ 27 l^5 D_s^2 - 213 l^5 - 66 l^4 D_s^2 -
3 l^4  + 27 l^3 - 3 l^3 D_s^2 + 16 l^2 D_s^2 - 3
l^2 +
$$
$$
9 l + l D_s^2 + D_s^2) D_s ) /
((1+ 3 l^2)^3 (l+1) (-1+l^2))
$$
where
$$
l=\sqrt{a}
$$
and
$$
w_2 = w_{02} / ((l+1)^2 (-1+l^2) (1+3 l^2)^6 l^2
)
$$
$$
w_{02} =  \sum_{i=0}^3 q_{2i} D_s^{2i}
$$
$$
q_{0} = - \frac{1}{12} - \frac{1}{6} l +
\frac{243}{4} l^{16} - \frac{135}{2} l^7 - 90 l^8
- \frac{87}{2} l^6 + \frac{243}{2} l^{14} +
\frac{81}{2} l^{12}
$$
$$
- \frac{153}{2} l^{10} -
\frac{39}{2} l^5 -
\frac{67}{6} l^4
-\frac{225}{2} l^9 - \frac{3}{2} l^2 -
\frac{81}{2} l^{11} + \frac{243}{2} l^{13} -
\frac{243}{2} l^{15} - \frac{17}{6} l^3
$$
$$
q_2 =
\frac{5265}{4} l^{12} + \frac{51}{2} l^6 + 117
l^{10} + \frac{333}{2} l^{11} + \frac{31}{8} l^4
$$
$$
+ \frac{17}{2} l^5 + \frac{5049}{2} l^{14} -
\frac{15309}{2} l^{19} + 2079 l^{13}
$$
$$
+ 31 l^8 -
8748 l^{18} + 42 l^9  + \frac{24057}{2} l^{21} +
\frac{103}{2} l^7 + \frac{16767}{8} l^{20} + 6561
l^{22}
$$
$$
+ \frac{4617}{2} l^{15} - 3159 l^{16} -
8262 l^{17}
$$
$$
q_4 =
- \frac{4413}{2} l^{13} - \frac{477}{2} l^{15} +
7047 l^{21} - \frac{1161}{4} l^{18}
$$
$$
+\frac{361}{6} l^8 +\frac{105}{2} l^{11} -
\frac{104247}{4} l^{22} - \frac{3}{4} l^4 + 17091
l^{20}
$$
$$
- 5103 l^{23} + \frac{43}{6} l^5 -
\frac{31185}{4} l^{19} + 7965 l^{17} -
\frac{3213}{4} l^{12} - \frac{475}{3} l^{10} +
510 l^{14} - \frac{49}{6} l^7 + \frac{23}{4} l^6
$$
$$
+ \frac{1}{12} l^3 + \frac{44469}{4} l^{24} -
\frac{1}{12} l^2 - \frac{6903}{4} l^{16} +
\frac{74}{3} l^9
$$
$$
q_6 =
- 13122 l^{25} - \frac{40419}{8} l^{18} +
\frac{17253}{2} l^{20} - \frac{24057}{2} l^{22} +
4374 l^{24} + \frac{6561}{2} l^{26} -
$$
$$
\frac{48843}{2} l^{21} - \frac{623}{24} l^8 +
\frac{59}{36} l^6 - \frac{3609}{4} l^{14}
$$
$$
-\frac{57}{8} l^{12} + \frac{411}{4} l^{10} +
\frac{11}{24} l^4 + \frac{1}{72} l^2 -
\frac{29}{12} l^7
-\frac{345}{4} l^{11} + \frac{1}{36} l^3  +
$$
$$
\frac{13}{36} l^5  + \frac{65}{4} l^9 +
\frac{2421}{4} l^{13} - \frac{7857}{4} l^{15} +
$$
$$
\frac{8667}{4} l^{17} + 28431 l^{23} + 7047
l^{19} + \frac{24705}{8} l^{16}
$$
Having integrated this expression we get
corrections to  droplets number.
The term at $s$
gives zero after integration and the first
non zero correction
has an order of $\sim s^2$
and doesn't depend on the total number of droplets.
A coefficient at $s^2$
has at $a=1/4$
a value
$$ d_0 =
311/3024
\ll 1
$$
At arbitrary $a$
a coefficient at $s^2$
in correction for the total number of droplets
will be
$$
d_a =
\frac{1}{72} [ 108 a^6 + 540 a^{11/2}
- 72 a^5 - 930 a^{9/2} - 336 a^4 +
$$
$$
713 a^{7/2} + 158 a^3 +4 a^2 - 203 a^{5/2}
- 6 a + 39 a^{3/2} - 3 a^{1/2} ] /
[ a^{3/2} (1-a)^2 (1+3a)]
$$

It will be interesting to compare results with
and without corrections from transition from
Poisson's distribution to Gauss
distribution. So, we
consider now this case.
At the leading term there will be no change. At
correction terms we
have
$$
y = -
\frac{17}{74088} D_s (-2331+289D_s^2)s +
(\frac{17}{196} D_s^2 +
$$
$$
\frac{732037}{12446784} D_s^4 +
\frac{24137569}{10978063488} D_s^6 -
\frac{13}{36}) s^2
$$
$$
d_0 =
-37/126
$$
$$
d_a =
\frac{1}{72}
( 648 a^{11/2}  - 216 a^5 - 1062 a^{9/2}
+108 a^4 + 753 a^{7/2} -
$$
$$
30 a^3 -
195 a^{5/2} - 12 a^2 +19 a^{3/2} + 6 a
- 7 a^{1/2} ) /
( a^{3/2} (1+3a) (1-a)^2 )
$$

It is seen that these corrections are small. At
arbitrary $a$ except too small ones
and those close to $1$ (these
values are unreal) we get values shown at
fig.14


\begin{figure}[hgh]

\includegraphics[angle=270,totalheight=10cm]{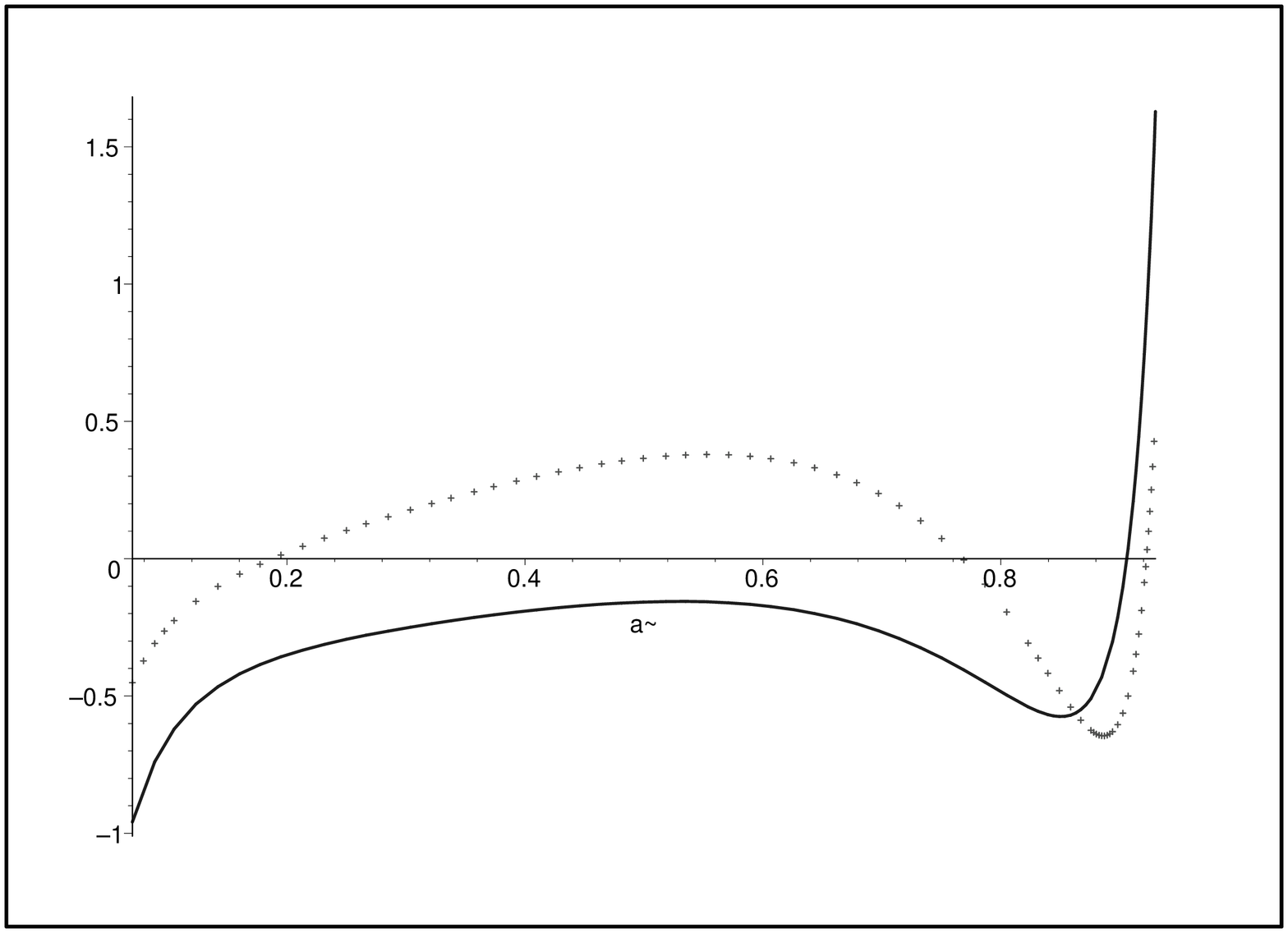}

\begin{caption}
{
Dependence of corrections, i.e. of
$d_a$
on $a$}
\end{caption}
\end{figure}

A point curve  shows corrections with transition
from Poisson's distribution to Gauss
distribution
taken into account,
a line  shows corrections without transition
from Poisson's distribution to Gauss
distribution. Both corrections have one and the
same order and they are small.

We see the plateau
for all $a$ except $a \ll 1/4$ and $1-a
\ll 1/4$.
So, here the smallness of corrections due to the
interactions
of stochastic peaks is quite clear.

But still the small value of
corrections can lead through sequential
influence  to essential change  of the
total droplets  number.  Conclusion
about  the  smallness of correction in
the total number of droplets  can be
made on the base of specific  kinetic
features  of the process (see "the
effect of the first droplet") which
will be done later.

One has to note that we have not taken into
account corrections from transition from
summation to integration. This is
definitely required by  discrete
character of droplets number.
It is made due to reasons
formulated below.
Really, we have at transition from summation in
formula
$$
P(N) = \sum_{\hat{N_1}} P_1 (\hat{N_1}, N_1) P_2
(N-\hat{N}_1-\hat{N}_2, N_2(\hat{N}_1))
$$
to integration\footnote{Here $N$ is the total number of
droplets, $\hat{N}_1$ is the stochastic
number of droplets in the first group,
$\hat{N}_2$ is the stochastic
number of droplets in the second group,
${N}_1$ is the mean number of droplets in the
first group, ${N}_2$
is the mean number of droplets in the
first group, which is a function of $\hat{N}_1$.
The value $N$ remains stochastic number.}
$$
P(N) = \int d \hat{N_1} P_1 (\hat{N_1}, N_1) P_2
(N-\hat{N}_1-\hat{N}_2, N_2(\hat{N}_1))
$$
to use the Euler-McLorrain's  decomposition.
It brings to asymptotic serial,
which can be included into a final
answer.

This
is the formal solution of the problem.
But discrete character in nucleation isn't so
trivial. The process of vapor consumption
can not begin
without the first droplet.
The system will wait for droplet as long as it
will be necessary.
It shows that discrete effects are complicate and
require a separate publication. At least one has
to put the initial moment at the moment of the
first  droplet appearance and then to consider
condensation with the substance in the first
droplet calculated explicitly as $\sim z^3$

This property will be called as
"the effect of the first
droplet". Here it leads only to the small effects, but in
the situation of smooth variation of external
 conditions the effect can be
greater.

\subsection{Self similarity of Gaussians}

To use Poisson's distribution for the first group
of droplets
one has to make the following notation. Really
nucleation conditions for the first group don't
differ from the whole group. So, for distribution
for the first group one has to take distribution
$P_1$ with reduced half-width.
But one can not attribute a half-width to
Poisson's distribution. That's why we considered
effects with and without corrections from
transition from Poisson's to Gauss distribution.
So, we can use Gauss distributions as initial
ones.
For Gauss distribution one can easily
reconsider the half-width. Then for
$P_1$
one can take
$$
P_1 \sim \exp( - (\hat{N_1} - N_1)^2 / (2 b  N_1))
$$
where $b$
is a renormalization coefficient.
Distribution $P_2$ remains previous
$$
P_2 \sim
\exp( - (\hat{N_2} - N_2 )^2 / (2   N_2))
$$
where $N_2$
is given by
$$
N_2 = (1-\frac{1}{3} s + \frac{2}{9} s^2
- \frac{14}{81} s^3 + \frac{35}{243} s^4 -
\frac{91}{729} s^5 - a)N
$$
where
$N$ is a mean total number of droplets,
$$
s = \frac{\hat{N_1} - a N}{a N }
$$
is a small parameter of an order
$N^{-1/2}$

After integration one comes to
$$
P \sim \exp( -
\frac{9 a }
{2(9 a  + b - 6 b a + 9 b a^2  -
9 a^2)}d^2)
$$
where
$$
d= (\hat{N} - N)/\sqrt{N}
$$

The half-width of the distribution $P_1$
  must be equal to the half-width of $P$,
  which leads to
 \begin{equation} \label{opo}
 b= 9 \frac{a(1-a)}{-9 a^2 +15 a - 1}
 \end{equation}
This value is drawn in fig. 15.

\begin{figure}[hgh]

\includegraphics[angle=270,totalheight=10cm]{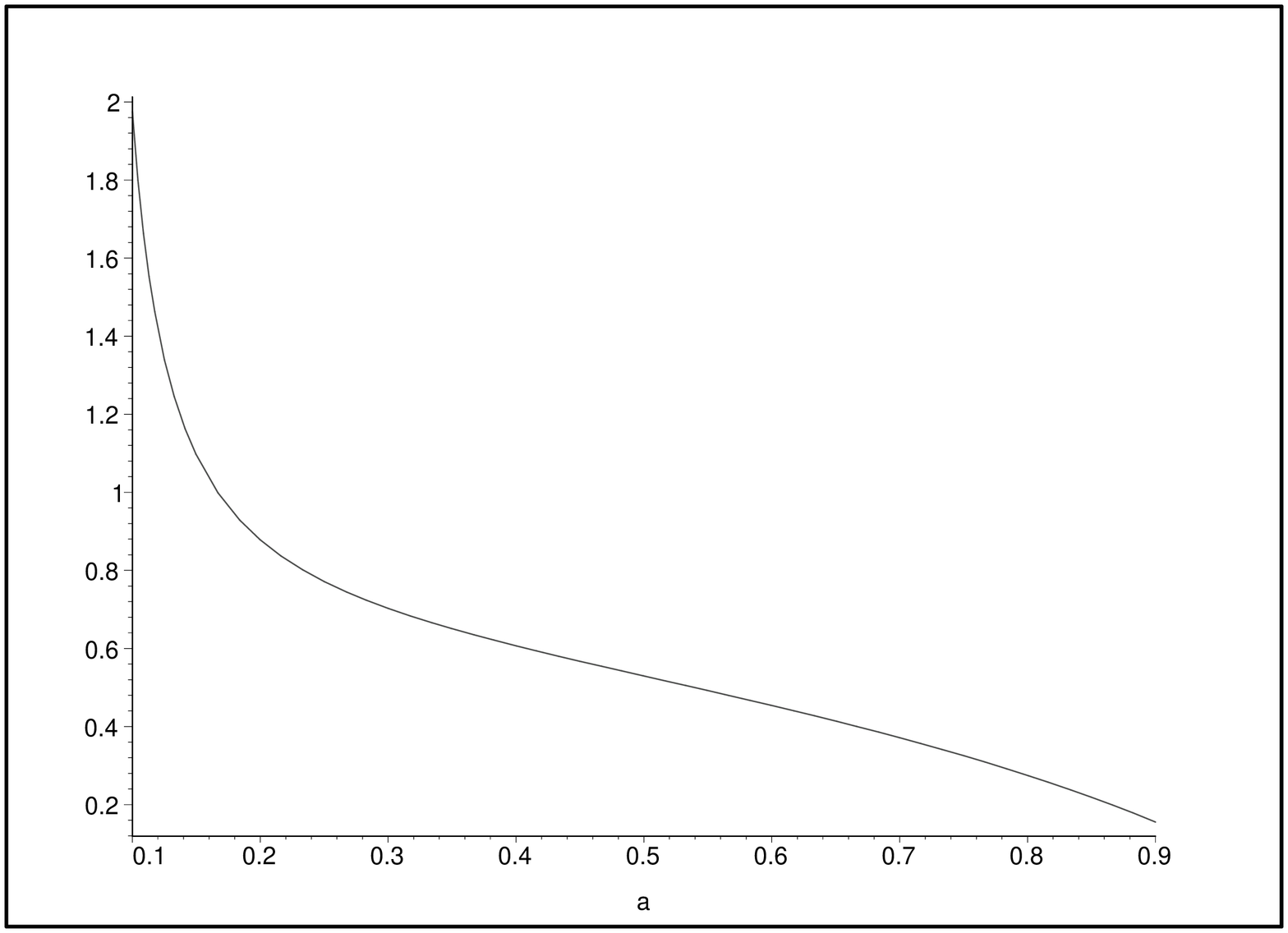}

\begin{caption}
{
Function $b(a)$.
}
\end{caption}

\end{figure}

Now we shall consider  effects of
renormalization. The ratio of corrections with
renormalization and without renormalizations is
given by
$$
r = \frac{1}{9}
\frac{\sqrt{-9 a^2 + 15 a -1}
\sqrt{3 a +1}}{a \sqrt{1-a}}
$$
and it is shown at fig.
16


\begin{figure}[hgh]

\includegraphics[angle=270,totalheight=10cm]{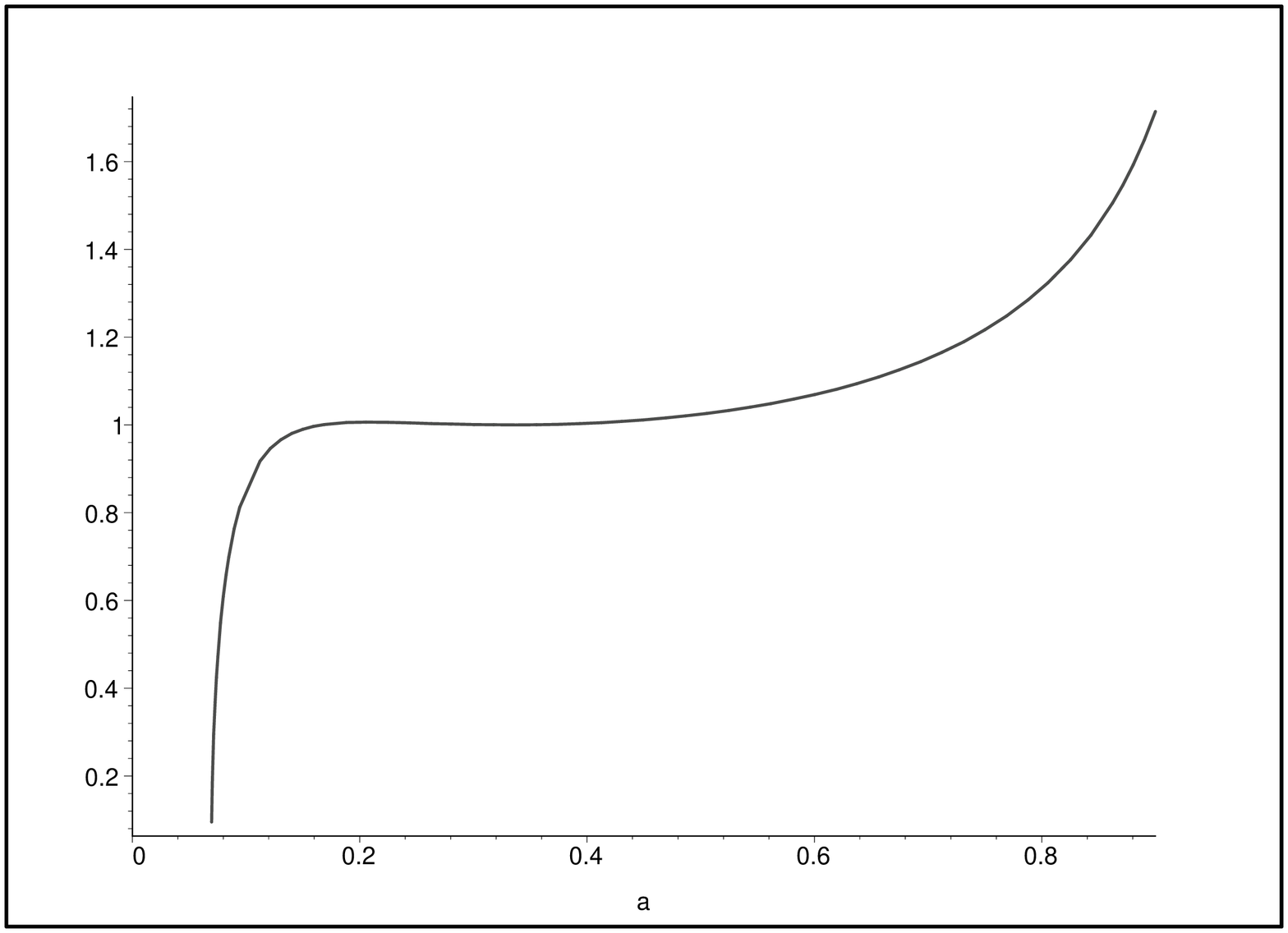}

\begin{caption}
{
Ratio of halfwidths
}
\end{caption}

\end{figure}

For all reasonable values of $a$
the last ratio  is approximately  $1$. At $a=1/4$
 we get
$1.0041$
Here the effect of similarity of nucleation
conditions doesn't lead to
remarkable effects in the change of half-width.
But here the smallness of change
takes place only because of
application of monodisperse approximation
and
later this change  will be essential.

Instead of taking into account all moments of
distribution we can directly calculate the effects
on the base of explicit form of spectrum in
frames of iteration procedure.

The result of
the previous consideration is the following: we
have proved that stochastic effects are small.
Beside this we have demonstrated how to use the
similarity of nucleation conditions.
Here it doesn't lead
to any remarkable effect,
but later
under the
smooth external conditions
this method will lead to some
essential numerical corrections.

One has to stress
that in \cite{Koll} a linearization
over deviation of the droplets number
from the mean value
was
made.
It was necessary to perform calculations. But in the
linear theory one can not get the deviation of the
mean value  of result
from the value calculated on the base of averaged
characteristics.
So, the attempt to get deviations in the
mean value of droplets in  \cite{Koll} is
senseless.

\subsection{
Calculation of dispersion}

Since the
mentioned combination  of gaussians is
also a gaussian
characterized by a
mean value and
by a dispersion one has to
determine these values for the
distribution of the total
number of droplets.
As it was shown above the mean value
of the droplets number is
practically the same as it is
prescribed by TAC and
we are interested in dispersion.
Now we shall calculate this value properly.

The most advanced approach to solve this
problem\footnote{Certainly, already  expression
(\ref{opo})  gives  the decrease of
the Gaussian halfwidth and can lead
directly to dispersion where $b$ is
smaller than a standard one. But $b$
essentially depends on $a$. For $0.2
\leq a \leq 0.9$ one can see the
approximate formula $b = 0.6 - (a-0.5)
\frac{0.7 - 0.2}{0.9 - 0.3}$
and
the absence of concrete value for $a$
doesn't allow to get a concrete value
for $b$.
 }
was suggested in
\cite{Koll}.
But even this approach has many disadvantages and
we need to reconsider it.

We shall characterize a droplet by
a linear size $\rho$
which is the cubic root
of its molecules number. Its velocity of growth
at fixed supersaturation
does not depend on $\rho$.

Decomposition of a whole interval of nucleation
into elementary intervals
is connected with some
difficulties.
An elementary length $\tilde{\Delta}$
according to \cite{Koll}
must satisfy
two requirements:

1. A number of droplets formed during elementary
length must be great.

2. An amplitude of a spectrum has to be
approximately constant during an elementary
interval.

It is clear that the second requirement
can not be satisfied. Stochastic deviations of an
amplitude leads to the
violation of the second requirement.

We shall apply the second requirement not to
the stochastic
amplitude as it was stated in \cite{Koll},
but to the averaged amplitude. Then the second
requirement is:

An averaged
amplitude of a spectrum has to be
approximately constant during an elementary
interval.

Now we shall see the evident
illegal consequences  of the
approach from \cite{Koll}.
"Stochastic" amplitudes  $f_i$ are
introduced in \cite{Koll} as
$$
f_i = \frac{N_i}{\tilde{\Delta}}
$$
where
$N_i$
is the number of droplets
formed during
$\tilde{\Delta}$.
It isn't the height of spectrum but
simply the renormalized
value of the droplets number appeared
during this interval.
An expression for the number of molecules
in droplets formed during interval number $i$
at a moment  $t_k $ (or $z_k$)
(it means that now we are at
interval number $k$)
with approximately constant rate of
nucleation $\bar{f}_i$
would be the following one
$$
\int_{x_{k-i}}^{x_{k-i+1}} \bar{f}_i \rho^3 d \rho =
\frac{1}{4} \bar{f}_i (x_{k-i+1}^4 - x_{k-i}^4)
$$
Namely, this equation was
derived in \cite{Koll} and
forms the base of further consideration.
Here $x_{k-i}$ is the
coordinate $\rho$ of the droplet which was
born at $z_i$ at a
moment $z_{k}$  (it corresponds to the
definition $x=z-\rho$)
The difference between forth powers corresponds
to a constant amplitude of spectrum. It is wrong
and then eq. (12) in \cite{Koll} and all further
equations  are not correct.

But the is no necessity to use such a way to
make an
account  of the number of molecules in a new phase.
It is absolutely sufficient to take the following
expression
$$
\int_{x_{k-i}}^{x_{k-i+1}} \bar{f}_i \rho^3 d \rho =
N_i  x_{k-i}^3 \approx N_i x_{k-i+1}^3
$$
which is valid at  $k-i \gg 1$.
In a whole quantity of substance it is sufficient
to take into account only droplets with $k-i \gg 1$.
The relative weight of dismissed terms will be
small.

Then for the
total number of molecules in droplets at
interval number $k$ we have the following
expression
$$
Q_k = \sum_{i=1}^k N_i x_{k-i}^3
$$
where $x_{k-i}$ is a corresponding
coordinate. This expression
can be rewritten as
$$
Q_k \sim \sum_{i=1}^k N_i  \hat{\Delta}^3 (k-i)^3
$$

This representation is important because now the
note  in
\cite{Koll}
after eq. (15)
isn't necessary. That note stated that the
probability for
$N_i$
to deviate from   the number  $\bar{N_i}$
of droplets calculated
in TAC under  the supersaturation
formed by stochastically
appeared droplets  in previous
intervals
is very low. That note
is doubtful because namely these deviations are
the base for stochastic effects. Now there is no
need in this note.

The next step is to build \cite{Koll} a
two cycle construction for nucleation period.
During  the first cycle
the main consumers of vapor appeared
in a system and during
the second cycle  they govern a process of
formation of all other droplets. In
\cite{Koll}
it is supposed that during the first cycle a
vapor depletion is negligible and during the
second cycle new droplets are absolutely governed
by droplets from the first cycle.  Now we shall
analyze an effectiveness of such procedure.

In TAC the corresponding evolution
equation will be
$$
\psi  (z) = \int_0^z (z-x)^3  \exp(-\psi(x)) dx
$$
The first iteration \cite{PhysRevE94} is
practically a precise solution and it gives the
number of droplets
$$
N_{tot} =
\frac{1}{4} \frac{4^{1/4} \pi \sqrt{2}}{\Gamma
(3/4)} = 1.2818
$$

A model solution requires that until  $z=p$
there will be no depletion of vapor and
then only the droplets formed before $z=p$
will consume vapor. Then for a total number of
droplets we have an expression
$$
N_{tot \ appr} = p + \int_p^{\infty}
\exp(-\frac{1}{4} x^4  + \frac{1}{4} (x-p)^4 ) dx
$$
A ratio $ q = N_{tot \ appr}/ N_{tot}$
is given in fig. 17

\begin{figure}[hgh]

\includegraphics[angle=270,totalheight=10cm]{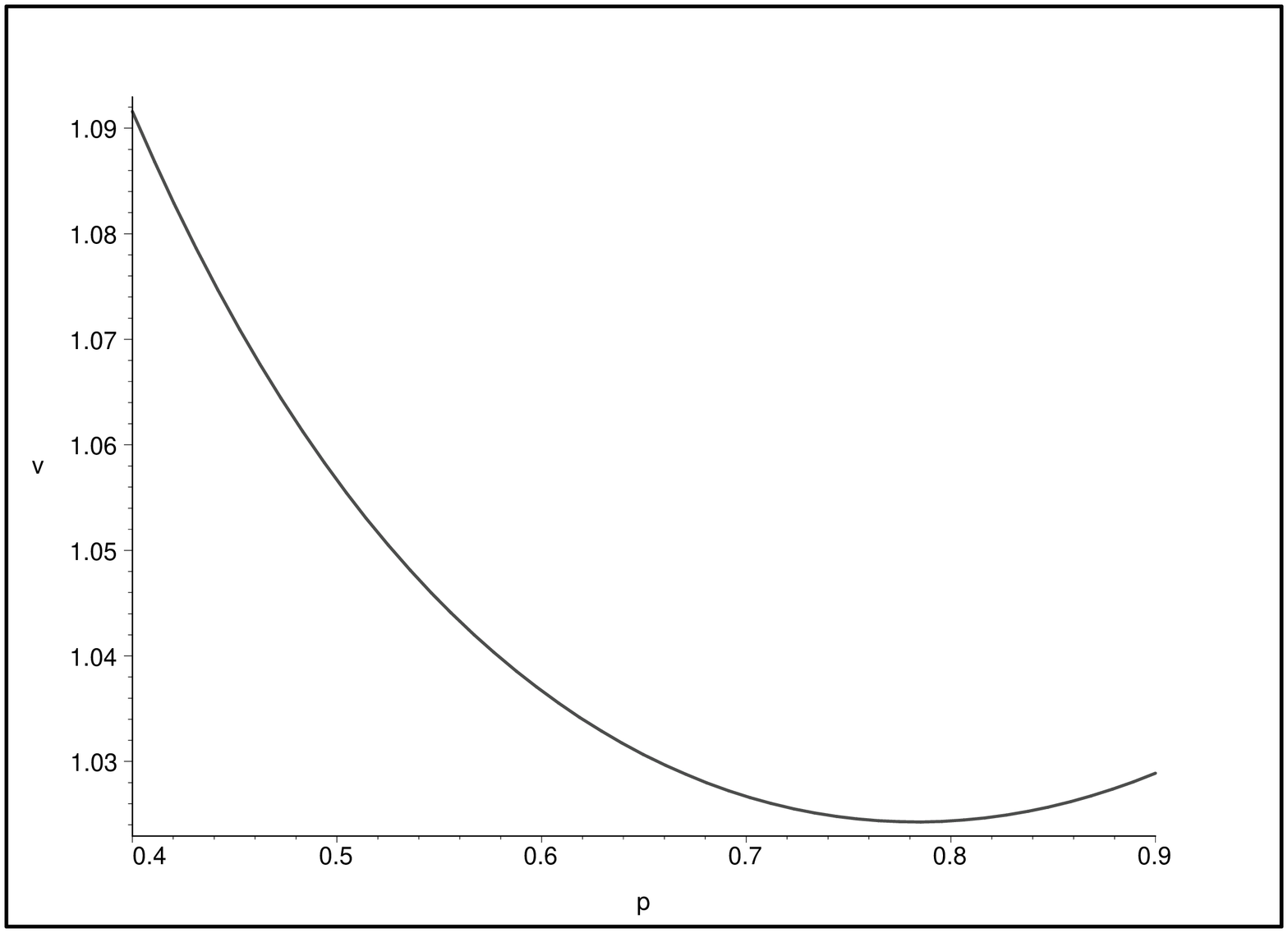}

\begin{caption}
{
The ratio of mean numbers of droplets }
\end{caption}

\end{figure}

Always $N_{tot\ appr}$ is greater than $N_{tot}$.
The value of
minimum corresponds to $p \sim 0.78$ which is
$55$ percent of the
total length of spectrum.
We can  stress the smooth dependence
$N_{tot \ appr}$ on $p$.

It is clear that in \cite{Koll}
the value of parameter of separation into two
cycles was not chosen in a good style (at least
from the point of view of TAC).
It corresponds to $p=0.6 4$.

Now we shall study the probability $P_k$
of formation  of stochastic number  $N_k$
of droplets at the first  $k$
 elementary intervals.
Our constructions now resemble  \cite{Koll}
but there is one
essential difference. We have no necessity to
linearize expression with respect to
$(N_i - \bar{N_i} )/ \bar{ N_i}$,
where
$N_i$
is a stochastic
number of droplets formed at interval
$i$,
$\bar{N_i}$ is a mean number of droplets formed
at interval $i$ (it is a function of stochastic
numbers of droplets at preceding intervals).
This linearization can not take place because
a  ratio $(N_i - \bar{N_i}) / \bar{ N_i}$
can be zero or can attain huge value (with  a low
probability).
It is more simple and more justified to linearize
expression on
$\sum_i \rho_i^3 (N_i - \bar{N_i}) / \bar{ N_i}   $
where $\rho_i$
is a linear size of droplets formed at interval $i$
(all of them have approximately the same size).
Really, due to summation the relative variations of
$\sum_i \rho_i^3 (N_i - \bar{N_i}) / \bar{ N_i}   $
are much smaller than variations of
 $(N_i - \bar{N_i}) / \bar{ N_i}$.

Variations of $(N_i - \bar{N_i}) / \bar{ N_i}$
would be small only at very big numbers of droplets
$N_{tot}$.
 One can get
$$
(N_i - \bar{N_i}) /
\bar{ N_i} \sim \bar{N_i}^{-1/2} $$
$$ N_{tot} \sim M \bar{N_i}$$
$M$ is a number of elementary intervals.
So, the  theory with linearization  proposed in
\cite{Koll}
would be well justified only in a region where
the result can be obtained on the base of
averaged characteristics.
The internal contradiction between the big
number $M$  and
the smallness of
fluctuations in the elementary interval
in \cite{Koll} is
obvious.

The linearization
proposed here is much more weak than in \cite{Koll}.
But it leads to the analogous numerical
expressions as in \cite{Koll}.
So, restrictions from \cite{Koll} are not necessary.

For dispersion of the total
distribution the result proposed in
\cite{Koll} was the following
$$
D^{\infty} = \tilde{N^{\infty}}
(1-\frac{\beta}{\alpha})
$$
where
$$
\beta = \beta_1 - \beta_2
$$
$$
\beta_1 =
8
\int_{1/2}^{\infty} d \xi \int_0^{1/2} d \tau (\xi -
\tau)^3
\exp(-\xi^4)
$$
$$
\beta_2 =
16
\int_{1/2}^{\infty} d \xi
\int_{1/2}^{\infty} d \eta
\int_0^{1/2} d \tau
(\tau- \xi)^3
(\tau-\eta)^3
 \exp(-\xi^4)
 \exp(-\eta^4)
$$
$$
\alpha = \int_0^{\infty} dx \exp(-x^4)
$$

In  the two-cycles
construction the value of $\alpha$,
which is proportional to the total number of
droplets has to be reconsidered and recalculated
on the base of two cycles. Then we have to use
instead of previous
$\alpha
\equiv \alpha_0$
a new value
$$
\alpha = \alpha_1 \equiv
1/2 + \int_{1/2}^{\infty}
\exp(-x^4 + (x-1/2)^4) dx
$$

In our approach we shall use parameter $k$
of separation\footnote{Here for simplicity we
use $k$ instead of $p/4^{1/4}$.}
of two cycles and we shall
calculate
$\alpha_1$ as
$$
\alpha_1 \equiv
k + \int_{k}^{\infty}
\exp(-x^4 + (x-k)^4) dx
$$
Then according to fig. 17 we see that
 the ratio    $\alpha_0 / \alpha_1$ is greater
 than $1$ and
  $\alpha_1$ is greater than $\alpha_0$.
Here we see that
two-cycles construction is approximate one.
Then result for  $D^{\infty}$
will differ from the number published in
\cite{Koll}
and will be
(here one has to put $k=1/2$)
$$
D^{\infty}_e = \tilde{N^{\infty}}
0.69
$$
instead of
$$
D^{\infty}_f = \tilde{N^{\infty}}
0.67
$$
as it is written in \cite{Koll}.

Numerical simulations show
that the value $D_e^{\infty}$ is
 one tenth more than a real result.
 So, the new theory is
necessary.

Now it is necessary to decide what $k$ shall we
choose. At arbitrary $k$ the expression for $\beta$
will be the same but for $\beta_1$  è $\beta_2$
we have
$$
\beta_1 =
8
\int_{k}^{\infty} d \xi \int_0^{k} d \tau (\xi -
\tau)^3
\exp(-\xi^4)
$$
$$
\beta_2 =
16
\int_{k}^{\infty} d \xi
\int_{k}^{\infty} d \eta
\int_0^{k} d \tau
(\tau- \xi)^3
(\tau-\eta)^3
 \exp(-\xi^4)
 \exp(-\eta^4)
$$
We have also to reconsider expression for $\alpha$.

After calculations we
have for dispersion as function of
$k$ the following fig.
18

\begin{figure}[hgh]

\includegraphics[angle=270,totalheight=10cm]{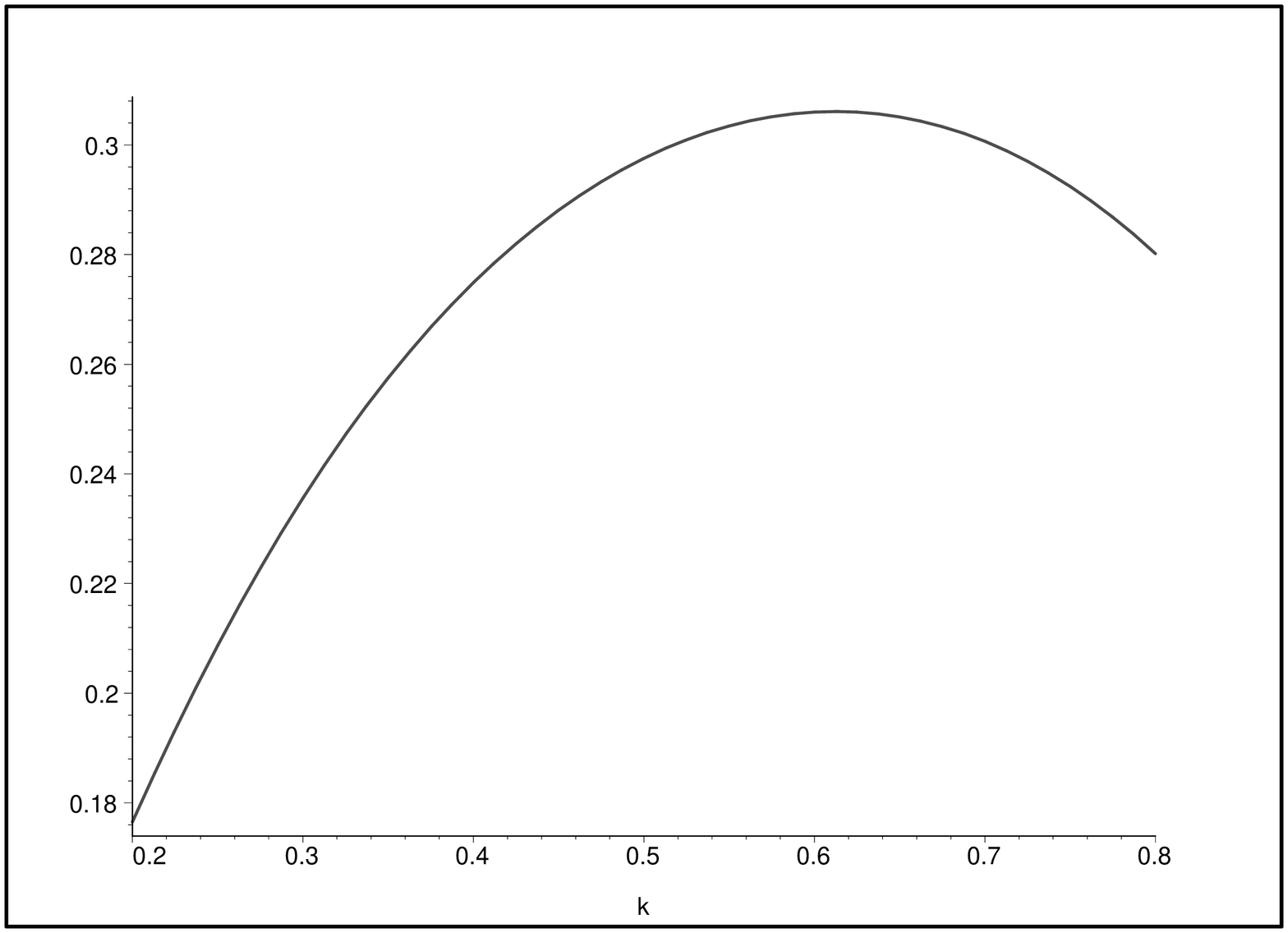}

\begin{caption}
{Relative deviation in dispersion as function of  $k$}
\end{caption}

\end{figure}

A minimal dispersion will be at $k=0.6$.
This value approximately equals
to $k=0.55$ which ensures  minimum of $\alpha_1$ where
the result is the
most close to the real value in the
number of droplets.
One can also add that this
 is the true value of $k$ because
 namely this value
corresponds to the sense of minimal work
when we have the low dispersion. It
corresponds to a minimal (in a certain sense)
entropy and
later we can get additional work from
increase of entropy.

Dispersion at $k=0.6$ will be
$$
D^{\infty} = 2 \tilde{N^{\infty}}
0.66
$$

Now we shall calculate the value of dispersion more
accurate.
Due to the similarity of nucleation the first
cycle doesn't differ from the whole period.
   Function
   $\beta$
for the first cycle will be
$$
\beta = \beta_1 - \beta_2
$$
$$
\beta_1 =
8
\int_{k_1}^{k} d \xi \int_0^{k_1} d \tau (\xi -
\tau)^3
\exp(-\xi^4)
$$
$$
\beta_2 =
16
\int_{k_1}^{k} d \xi
\int_{k_1}^{k} d \eta
\int_0^{k_1} d \tau
(\tau- \xi)^3
(\tau-\eta)^3
 \exp(-\xi^4)
 \exp(-\eta^4)
$$
Calculations for $k_1 = 0.6*0.6 = 0.36$ and  $k=0.6$
give $\beta = \beta ' \equiv 0.0255$.

In the most rough approximation
one has to add
$\beta '  = 0.0255$ to the previous value of
$\beta= 0.305$
which leads to  $\beta = 0.32\div 0.33$.
The smallness
of $\beta'$ in comparison with $\beta$ allows
to use this linear approximation.
For this value of $\beta$
the value of dispersion will be
$$
D^{\infty} = D^{\infty}_3 \equiv 2\tilde{N^{\infty}}
0.64
$$

It is interesting that
this result can be gotten by another
(rather artificial) procedure:

 We suppose
that
$\beta_1$ and $\beta_2$
are the first two terms of some series.
We don't know other terms, but
it is reasonable to suppose that the series
resembles geometric progression with denominator
$\beta_2 / \beta_1$.
This leads to dispersion
$$
D^{\infty} = 2 \tilde{N^{\infty}}
0.64107
$$
As it follows from fig.19 the   value of extremum
 remains
$k=0.6$.


\begin{figure}[hgh]

\includegraphics[angle=270,totalheight=10cm]{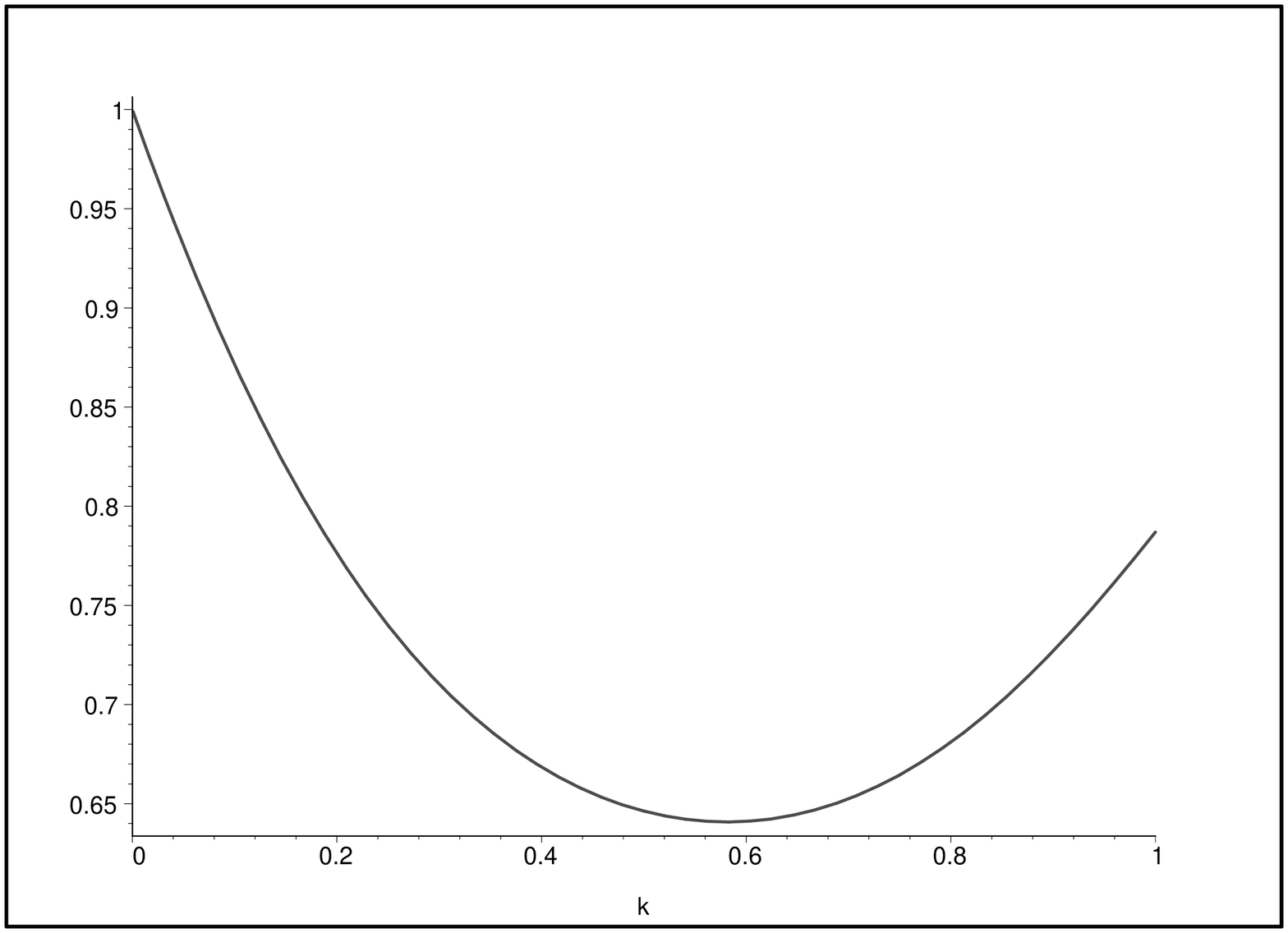}

\begin{caption}
{Dispersion.}
\end{caption}

\end{figure}

The
absence of the shift of extremum  is important and is
necessary for this approach to be a self consistent.

Now we shall see whether some other approaches  can
lead to essential reconsideration of result
for dispersion.

A way to make results more precise is
take into account the shift of dispersion
directly in initial formulas. Having written for
the dispersion correction in the first cycle
$$
D^{\infty}_3 = 2\tilde{N^{\infty}}
\gamma
$$
with parameter $\gamma$,
we get
for the final distribution
$$
P^{(k)}  \sim
\int_{-\infty}^{\infty}
dN_1 dN_2 ... d N_P
\prod_{i=1}^{P} \exp(-\frac{(N_i  - \bar{N_1})^2}{2
\gamma \bar{N_1}})
$$
$$
\exp[\frac{[N^{(k)} - \tilde{N^{(k)}} -
\sum_{j=1}^{P} a_j^{(k)} (N_j - \bar{N_1})]^2}
{2(\tilde{N^{(k)}}-P \bar{N_1})}]
$$
where
$P$
is the number of elementary intervals until
the argument
$k$,
$\tilde{N^{(k)}}$
is the number of droplets calculated on the base
of the theory with averaged characteristics,
$\bar{N_i}$
the mean number of droplets formed during
interval number
$i$
with account of
fluctuations from previous intervals.
The values $a_i^{(k)}$
 are given by
$$
a_i^{(k)} = 1 - \sum_{j=P+1}^{k}
\frac{\exp(-j^4/M^4)}{M^4} 4 (j-i)^3
$$
and $M$
is the total number of intervals.

Having fulfilled integration
$\int_{-\infty}^{\infty}
dN_1 dN_2 ... d N_P$,
we get for a limit value of dispersion
$$
D^{\infty} =  2 \tilde{N^{\infty}} ( 1-
\frac{ k (1-\gamma)}{\alpha} - \frac{\gamma
\beta}{\alpha})
$$

 The approximate similarity of spectrums
 leads to equation on $\gamma$,
which can be easily solved
 $$
\gamma =
\frac{1 - \frac{k}{\alpha}}
{1+ \frac{\beta}{\alpha}- \frac{k}{\alpha}}
$$
Calculations lead to
$$
\gamma (k=0.6) = 0.51
$$

This result is very strange. It radically differs
from the previous one.
Certainly we made an error.
The reason of the error in previous approach
is
that the duration of the first cycle is limited
by
$k$.
So, we have to limit the duration of a whole
period. The limit of the
whole nucleation is, evidently,
 $\sim 1 $.

The limit of integration
corresponds to the current moment
of time.
It can not be greater than $1$.
So, we have to take it equal
to $1$.
Then we have to recalculate $\beta$
as
$$
\beta_{initial} = \beta_1 - \beta_2
$$
$$
\beta_1 =
8
\int_{k}^{1} d \xi \int_0^{k} d \tau (\xi -
\tau)^3
\exp(-\xi^4)
$$
$$
\beta_2 =
16
\int_{k}^{1} d \xi
\int_{k}^{1} d \eta
\int_0^{k} d \tau
(\tau- \xi)^3
(\tau-\eta)^3
 \exp(-\xi^4)
 \exp(-\eta^4)
$$

To calculate the value of dispersion we can act
in two manners.
The first way is to calculate dispersion at
the point of extremum of $\beta$.
Now we shall show a dependence  $\beta_{initial}$
on
 $k$ at fig. 20.


\begin{figure}[hgh]

\includegraphics[angle=270,totalheight=10cm]{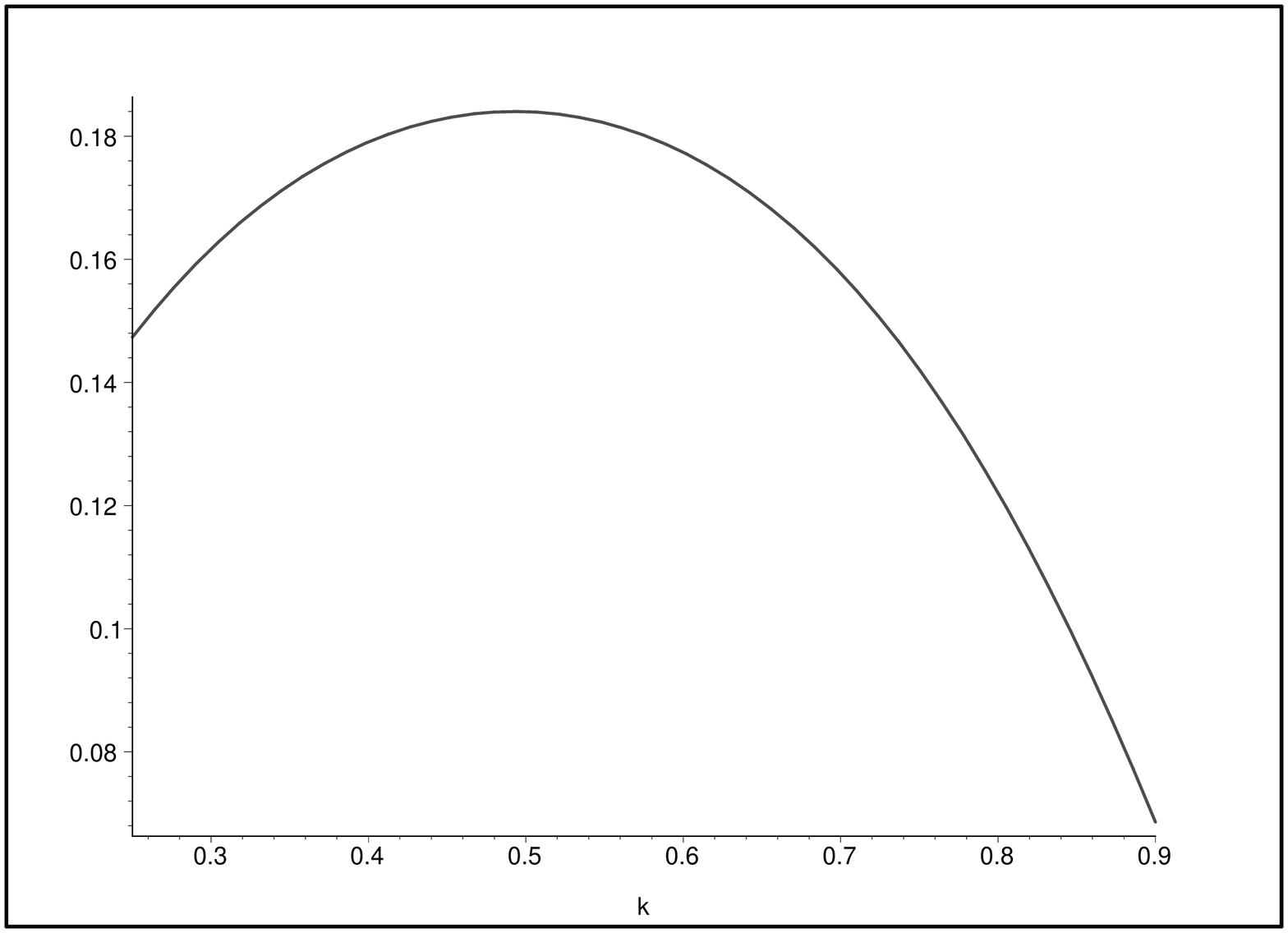}

\begin{caption}
{ $\beta_{initial}$ as a function of $ k$}
\end{caption}

\end{figure}

Calculations give $\beta_{initial} = 0.18$
and for the final dispersion
$$
D^{\infty}_3 = 2\tilde{N^{\infty}}
0.65
$$

This value practically
coincides with a previous approach.

It is necessary to stress that
one can not directly use extremal
properties of $\gamma$ to  get $p$
which provides $\beta$
extremum.
In reality the final characteristic is $\gamma$,
but the calculations
show that $\gamma$ has no extremum.
Certainly, this
is the weak point of approach based on
extremal properties.

Another way is to
use $k$ which provides the most precise
value for the
total number of droplets in this model.
Having compared the mean total
number of droplets in TAC
calculated in this
model one can see that this number has
minimum  $0.927$ which
is greater than the precise value
and the value $0.90$
given in the first iteration. This
extremum is
attained at $k=0.55$. Namely this value will be
chosen as $k$ and this leads to
$$
D^{\infty}_4 = 0.62
$$
which lies in frames of
precision of numerical simulation.

One can add that there
is no need to use $k$ corresponding to
extremum of the number of
droplets calculated up to $1$
instead of $\infty$. Then the minimum is $0.852$
and it is attained at
$k=0.45$. Actually
there is no reasons to take this value
because here the value
of minimum is strongly less than the
precise value and it
corresponds to the maximum of
deviation from
the precise value. So, there are no reasons
to take this value.

One has to stress
that the procedure adopted here is really
necessary. The two-period
model with a fixed boundary in
time scale used in \cite{Koll}
can not be the base of the correct
calculation of dispersion.
The reasons are the following
\begin{itemize}
\item
The behavior of supersaturation here is the model
behavior (MB) with parameters, which characterize the
influence of the first part of spectrum. In
investigation of stochastic effects the
fluctuations of these parameters are the source of
fluctuations of the total number of droplets. The
fuctional form of MB is chosen to ensure the
correct number of droplets in frames of TAC. The
change of parameters in MB leads to the imaginary
change of external conditions in TAC which
corresponds to the MB with given parameters.
\item
The change of external
parameters in TAC has to lead to the
change of the boundary between parts.
But this can not be
done in the model with  a fixed boundary.
\end{itemize}

In our approach the
summation of geometric progression and
all other
approaches ensure the possibility to overcome the
restrictions of the two parts
approach with a fixed boundary.
The summation of geometric progression describes the
equivalence of all points which is prescribed by the
property of similarity of spectrums.

The
relative
smallness
of numerical errors in \cite{Koll} is caused
by the following reasons
\begin{itemize}
\item
The moderate possibility of linearization (the
non-linearity isn't too big) in kinetics of decay
\item
The weak dependence $N_{tot}$ on $p$ in frames
of TAC
\item
The weak dependence of $N_{tot}$ on $N(p)$
in frames of TAC
\item
The effects of
stabilization are very strong. Really,
instead of
addition $0.5$ the benefit of the second part
to dispersion is only $0.64 - 0.5 = 0.14$
(or even less if we use $p\sim
0.55$
instead of $p \sim 0.5$)
\item
The existence of the special
buffer part of nucleation
period which will be described
later
\end{itemize}

\subsection{Numerical results}

Numerical simulation of nucleation can be done by
the following method. We split the nucleation
interval into many parts (steps). At every step a
droplet will be formed or not. The probability to
appear must be rather low, then we ensure  the
smallness of probability to have two droplets at
the same interval.  This means that the interval
is "elementary".

The process of formation is simulated by a
random generator in a range $[ 0, 1]$. If a
generated number is smaller than a threshold
parameter  $u$,
then there
will be no formation of a droplet. If it is
greater than a threshold, we shall form a droplet.
As a result we have spectrum $\hat{f} $
of droplets sizes.
Now it is a chain of $0$ and $1$.
The parameter $u$
descends according to macroscopic law
\cite{PhysRevE94}
$$
u = u_0 \exp(- \Gamma G /\Phi )
$$
from a theory with averaged characteristics (it
is based only on a
conservation law without any averaging
 and can be used).
Here
\begin{equation}\label{Gamma}
\Gamma \sim \frac{d F_c}{d \Phi} \sim \nu_c
\end{equation}
$
\Phi
$
is the initial supersaturation,
$F_c$
is a free energy of critical embryos formation,
 $\nu_c$
is a number of molecules inside a critical
embryo,
$G$
is the number of molecules in a new phase taken in
units of
a molecules number density in a saturated vapor.
By renormalization one can take away all parameters
except $G$.

To simplify calculations radically
one can use the following representation
\cite{PhysRevE94} for
$G$:
$$
G = z^3 G_0 - 3  z^2 G_1 + 3 z G_2 - G_3
$$
where   $z$
is a coordinate of a front of spectrum,
and $G_i$
are given by
$$
G_i = \int_0^z \hat{f}(x) x^i dx
$$
We needn't to recalculate $G_i$,
but can only ascend the region of integration,
having added to integrals the functions
$z^i\hat{f}(z)  dx$ at every step.

Our results are given below.
The interval is split into
$30000$
parts. Parameter
 $u_0$
have been varied from  $0$ up to $1$
which leads to a different
number of droplets. It is clear that the limit
values are not good: at  $0$
there are no droplets in the system, at $1$
our intervals are not elementary.
At every $u_0$
results were averaged over $1000$ attempts.

Shifts of droplets numbers are
drawn at fig.21  as a function of
$\ln \tilde{N^{(\infty)}}$

\begin{figure}[hgh]

\includegraphics[angle=270,totalheight=10cm]{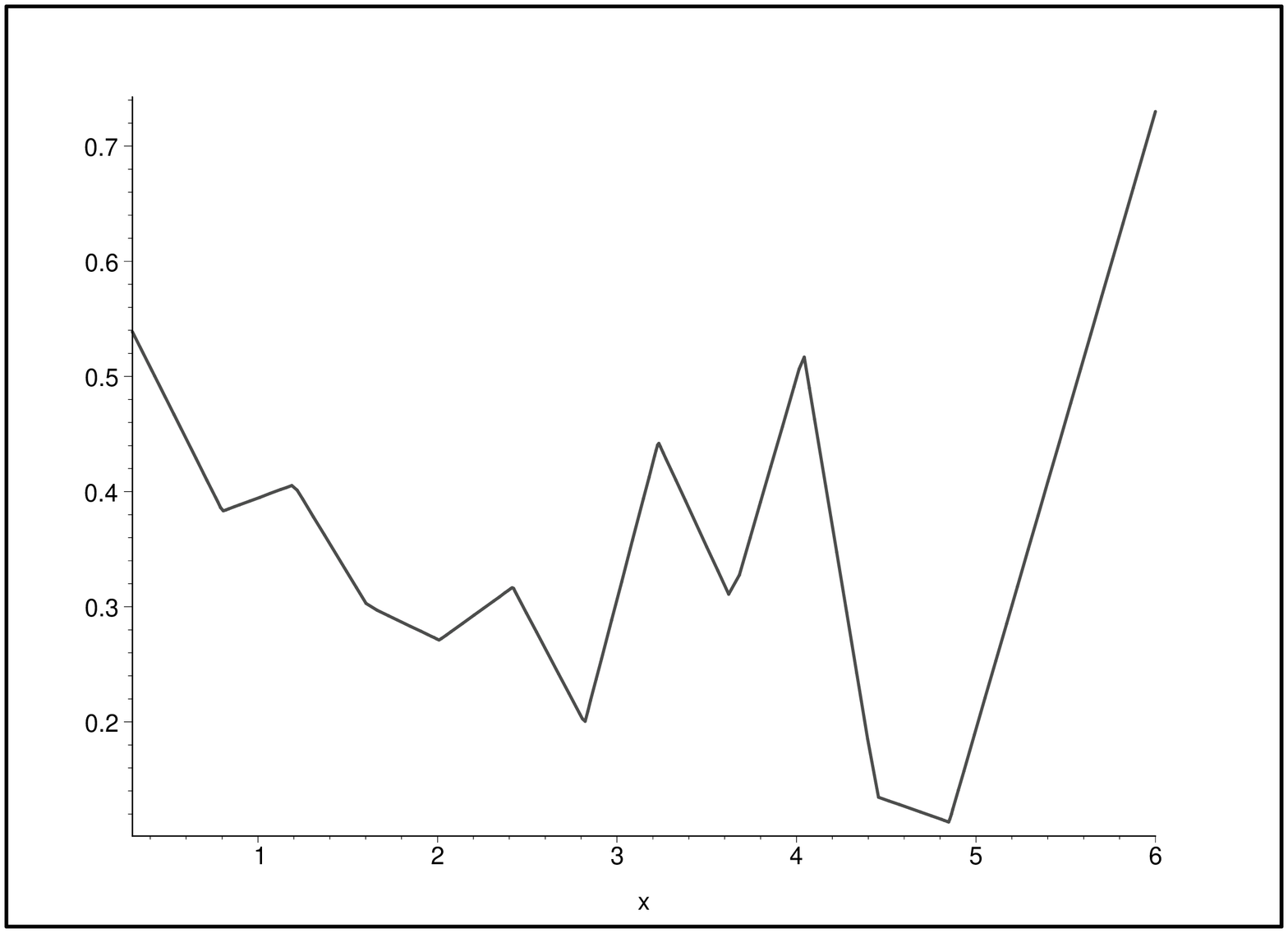}

\begin{caption}
{
Shifts of droplets numbers as a function of
 $\ln \tilde{N^{(\infty)}}$}
\end{caption}

\end{figure}
It is seen that an analytical result about
negligible value of corrections is correct.

Dispersion as a function of $\ln
\tilde{N^{(\infty)}}$
is shown in fig. 22.


\begin{figure}[hgh]

\includegraphics[angle=270,totalheight=10cm]{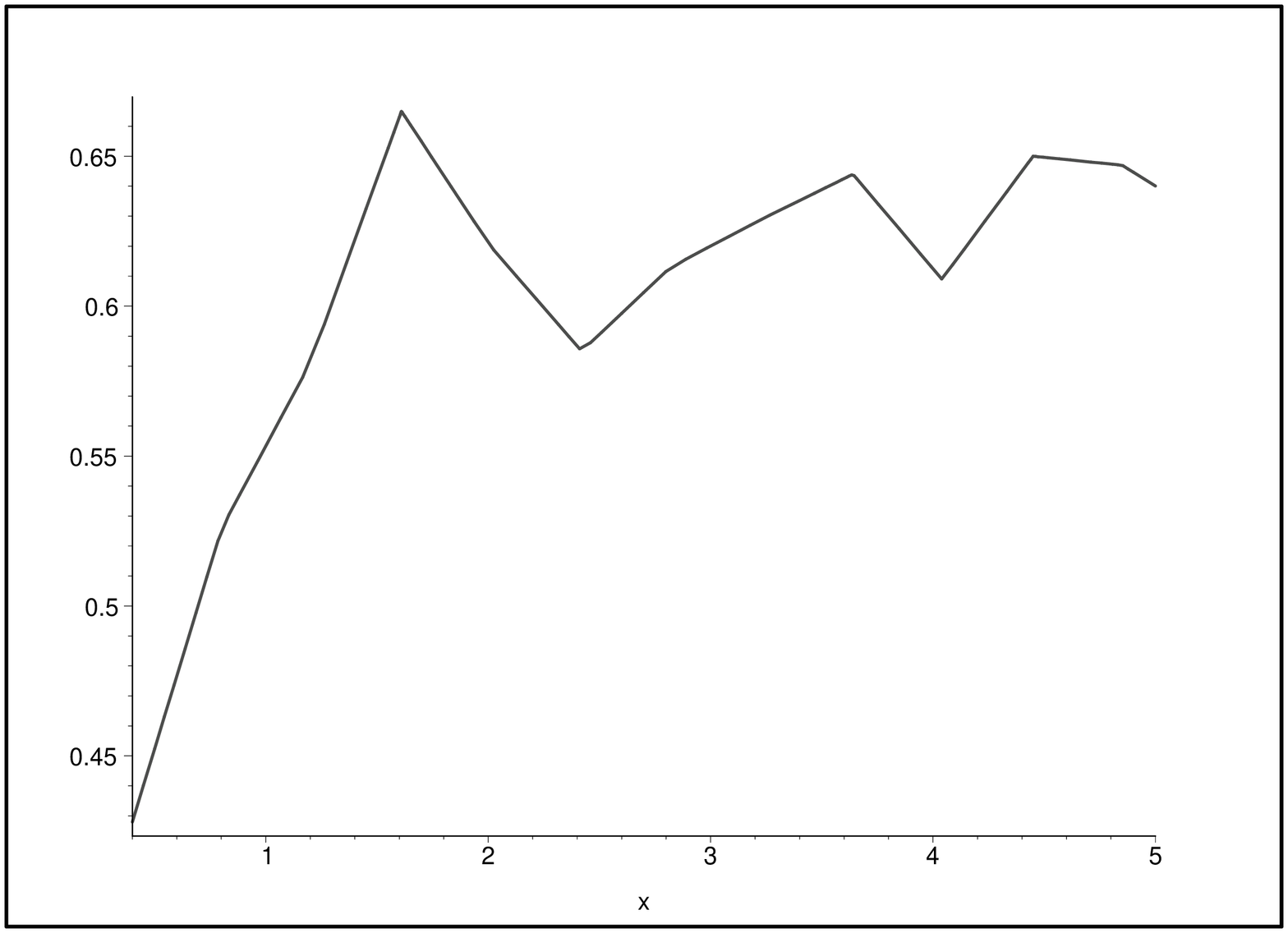}

\begin{caption}
{  Relative dispersion $\gamma$
as a function of  $\ln \tilde{N^{(\infty)}}$}
\end{caption}

\end{figure}

It is seen that  the analytical value of dispersion
coincides with numerical simulation.
The ends of the curve correspond to a zero
number of droplets and to a giant number of
droplets when the elementary intervals are not
elementary  and have to be thrown out.

Stochastic effects in dynamic conditions
\cite{PhysRevE94} can be
analyzed by the same method. We needn't to
describe it here. Numerical results are drawn
below. Fig. 23 shows the shift in the number of
droplets. It is small.
  Dispersion is drawn in fig. 24 (i.e. the value
  of
$\gamma$).
It is greater than in the case of decay.

\begin{figure}[hgh]

\includegraphics[angle=270,totalheight=10cm]{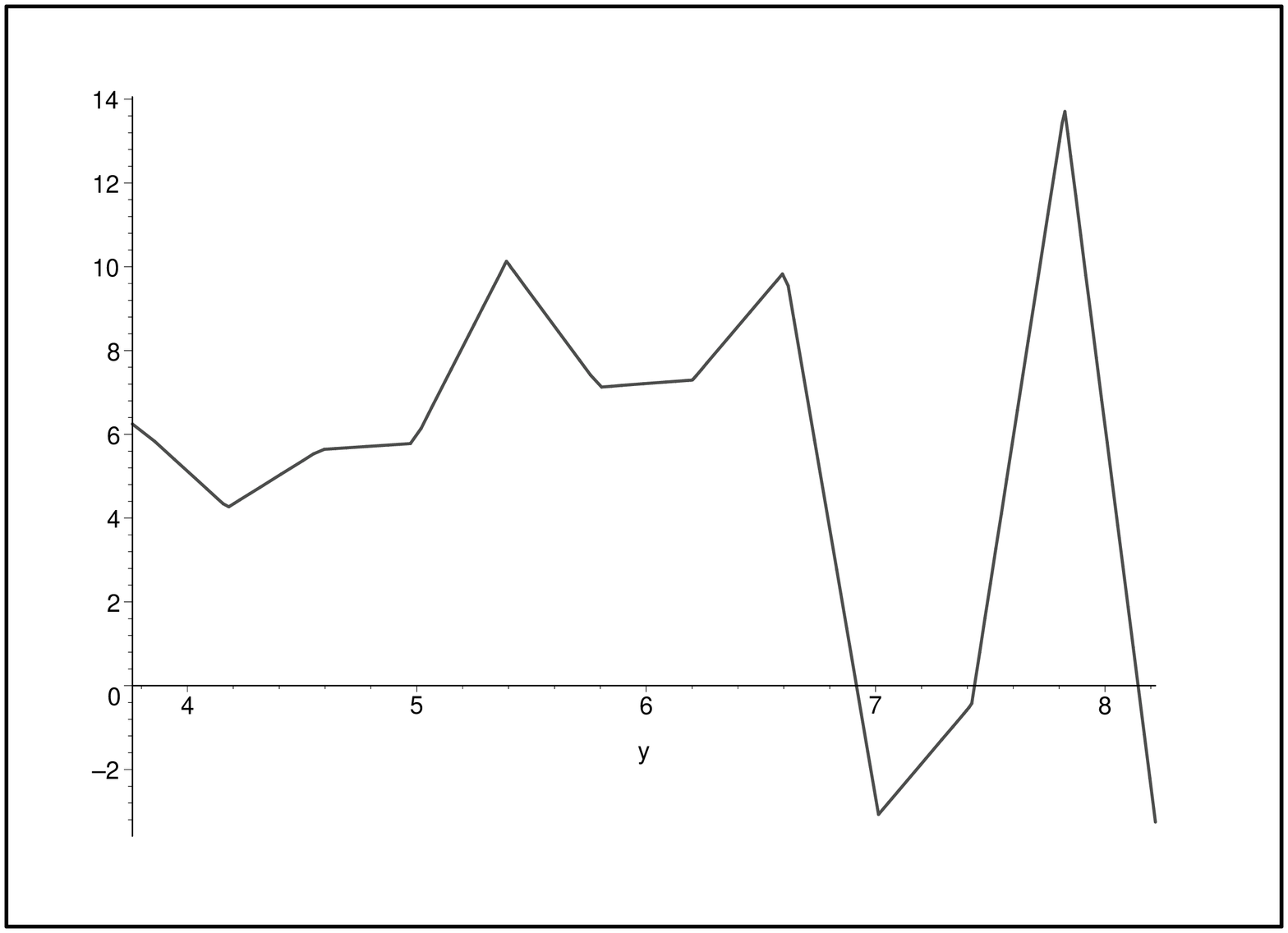}

\begin{caption}
{
Shift in a droplets number as a function of
 $\ln \tilde{N^{(\infty)}}$ for dynamic conditions}
\end{caption}

\end{figure}


\begin{figure}[hgh]

\includegraphics[angle=270,totalheight=10cm]{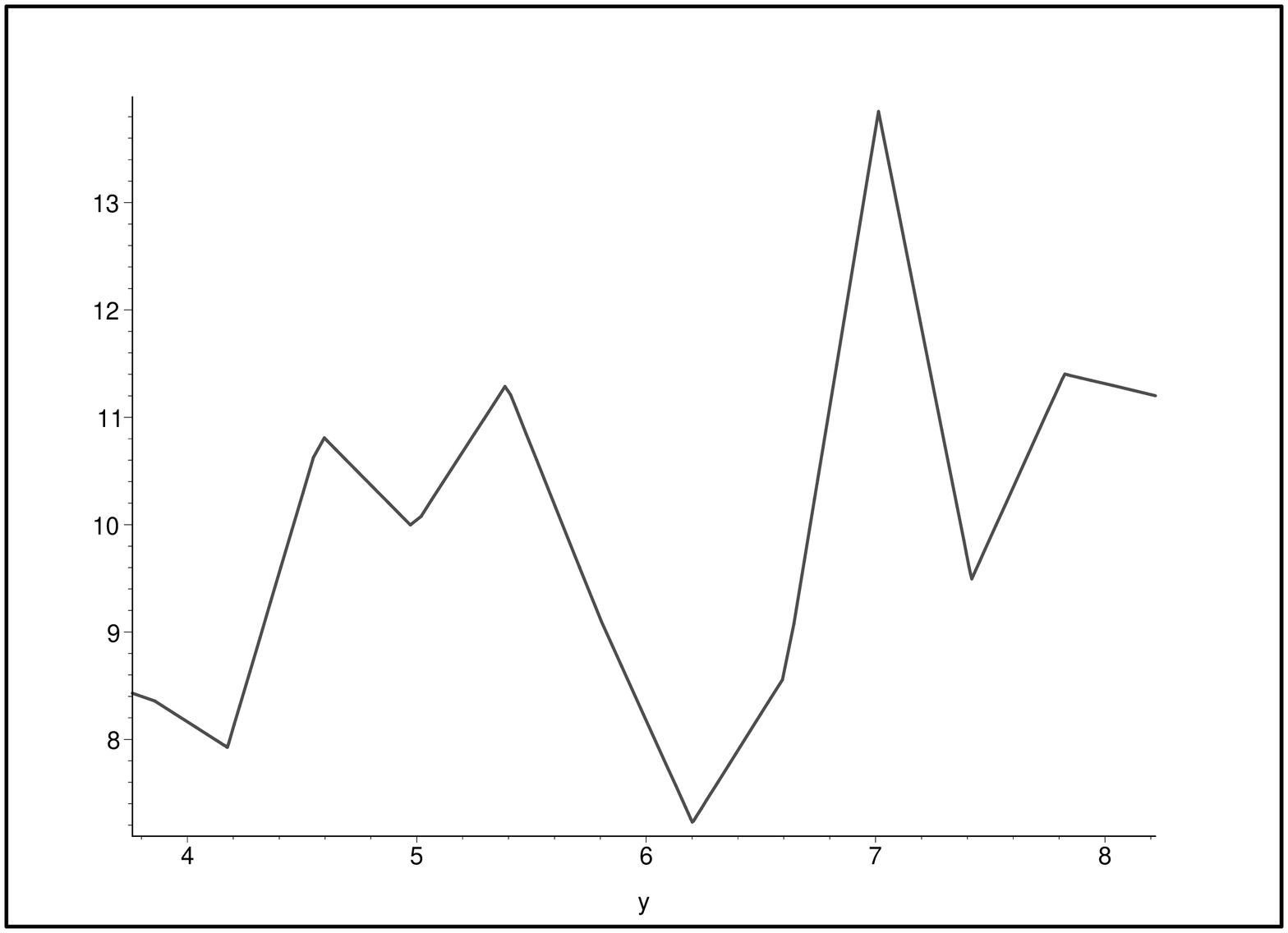}

\begin{caption}
{ Relative  dispersion $\gamma$
as a function of  $\ln \tilde{N^{(\infty)}}$
for dynamic conditions}
\end{caption}

\end{figure}

The physical reasons for the smallness of the
droplet number shift for decay and for dynamic
conditions will be different.

For decay the reason is the following. The system
wait the first droplet as long as necessary.
Actually the time for kinetics of
this system is $t(G)$ with
no connection with real time (certainly, the rate
of nucleation has such connection).
This phenomena is the reason for a smallness.

In dynamic conditions there is a time dependent
parameter - the intensity of external source. So,
 there is no such a reason.

But here in the theory with averaged
characteristics there is a property of a weak
dependence
of the total number of droplets
on microscopic corrections for a free
energy \cite{PhysRevE94}. The same is valid also
for fluctuation deviations. So there will be a
weak effect of stochastic nucleation.

Because the reasons for smallness of effect in
decay and dynamic conditions are different it is
interesting to see whether they continue to act
when
the supersaturation is stabilized at some moment.
Analytical results shows that the will be an
overlapping of two reasons.


Really, if stabilization
takes place at the period where the main
consumers of vapor are going to appear then the
majority of droplets appear in the situation when
there is no influence on the system. Then the
situation for these droplets resembles decay
conditions (and may be even better because the
external supersaturation \cite{book1} is going to
decrease). So the reason for the decay situation
works here.

If stabilization takes place at the
second cycle, then the behavior of
supersaturation is governed by droplets formed in
dynamic conditions and we have here the reason
for smallness in dynamic conditions. In both
situations the effect is small.
Numerical results
confirm this conclusion.

\subsection{Conclusions}

The main result of this publication is a correct
definition of all main characteristics of
stochastic nucleation.  It is shown that the main
role in stochastic effects belongs to all
droplets, but not to the main consumers of vapor.
Only the property of the nucleation conditions
similarity allows us to solve the problem of
account of all influences during the nucleation
period.

When all disadvantages of \cite{Koll}, \cite{Vest}
are shown it is clear that these publications can not
be considered as a solid base for nucleation
investigation.

But why results obtained in  \cite{Koll},
\cite{Vest}
are so close numerically to real values?
The reason is that on a level of averaged
characteristics there is a universality of
nucleation process. So, the errors of
\cite{Koll}, \cite{Vest}
cannot lead to a qualitatively wrong results.

One has to stress that all corrections obtained
in this paper are also universal ones. Certainly,
they are some coefficients in decompositions and
the functional form of decomposition
is prescribed now.

There is also a
second specific reason for the smallness of
an error in
numerical values presented in \cite{Koll}.
The reason is the following
\begin{itemize}
\item
The process of nucleation can be split in three
sequential parts.
\item
The first part is the part where the
main consumers of
vapor were appeared.
Here the vapor depletion is small
\item
The second (buffer)
part is the part where conditions are
ideal,
the vapor depletion is small, but the droplets
formed in this part can not
attain big sizes even
at the end of the whole nucleation period
and, thus,  can not consume
enough vapor and they are not
the main consumers of vapor
\item
The third part is the part
where the depletion isn't small
and droplets appeared in
this part aren't the main consumers
of vapor in the nucleation period.
\item
All parts have the lengths
of one and the same scale.
\end{itemize}
Certainly this structure
was not declared in \cite{Koll}
which  made the
derivation in \cite{Koll} illegal.

The existence of the buffer
part is necessary to balance the
errors appeared from
the fact that the fluctuations leads to
to the
absence of applicability of
functional approximations for the
nucleation kinetics. More correctly
is to use the property
of the
"internal time of decay"
which will be done separately.

It seems that all effects considered here are
negligible. For simple systems it is really true.
But for systems with more complex
kinetic behavior these effects can be giant. One
of such systems is already described
theoretically and this description
will be presented soon in
a separate publication.

To compare results given
here with the previous approach
one can simply recall
that the real error is the error
in the relative deviation
of dispersion from the standard value.
In \cite{Koll} this error is more than one quarter.
Here the error  is practically absent.

In diffusion regime of
droplets growth one has to use another
approach based on
\cite{PhysicaA},
\cite{PhysRevE2001}. In  \cite{PhysRevE2001}
an explicit description of nucleation with
account of stochastic effects was constructed.

\end{document}